\documentclass[jcp,graphicx,amsmath,amssymb,reprint]{revtex4-1}
\pdfoutput=1

\usepackage[locale=US]{siunitx}
\usepackage{hyperref}
\usepackage[capitalise]{cleveref}
\usepackage{bm}
\usepackage[]{tikz}
\usetikzlibrary{calc}
\usepackage{mathptmx}

\newcommand{\round}[2]{\num[round-mode=places,round-precision=#1]{#2}}

\begin{document}
{\footnotesize\vspace{-2\baselineskip} The following article has been submitted to The Journal of Chemical Physics. After it is published, it will be found at \href{https://publishing.aip.org/resources/librarians/products/journals/}{Link}.\\}

\title{Adaptive-precision potentials for large-scale atomistic simulations}

\author{David Immel}
\affiliation{Jülich Supercomputing Centre (JSC), Institute for Advanced Simulation (IAS), Forschungszentrum Jülich, Jülich, Germany}

\author{Ralf Drautz}
\affiliation{Interdisciplinary Centre for Advanced Materials Simulations (ICAMS), Ruhr Universität Bochum, Bochum, Germany}

\author{Godehard Sutmann}
\email[]{g.sutmann@fz-juelich.de}
\affiliation{Jülich Supercomputing Centre (JSC), Institute for Advanced Simulation (IAS), Forschungszentrum Jülich, Jülich, Germany\\Interdisciplinary Centre for Advanced Materials Simulations (ICAMS), Ruhr Universität Bochum, Bochum, Germany}

\date{November 5, 2024}

\begin{abstract}
Large-scale atomistic simulations rely on interatomic potentials providing an efficient representation of atomic energies and forces.
Modern machine-learning (ML) potentials provide the most precise representation compared to electronic structure calculations while traditional potentials provide a less precise, but computationally much faster representation and thus allow simulations of larger systems.
We present a method to combine a traditional and a ML potential to a multi-resolution description, leading to an adaptive-precision potential with an optimum of performance and precision in large complex atomistic systems.
The required precision is determined per atom by a local structure analysis and updated automatically during simulation.
We use copper as demonstrator material with an embedded atom model as classical force field and an atomic cluster expansion (ACE) as ML potential, but in principle a broader class of potential combinations can be coupled by this method.
The approach is developed for the molecular-dynamics simulator LAMMPS and includes a load-balancer to prevent problems due to the atom dependent force-calculation times, which makes it suitable for large-scale atomistic simulations.
The developed adaptive-precision copper potential represents the ACE-forces and -energies with a precision of $\SI{10}{\milli\electronvolt/\angstrom}$ and $\SI{0}{\milli\electronvolt}$ for the precisely calculated atoms in a nanoindentation of 4 million atoms calculated for $\SI{100}{\pico\second}$ and shows a speedup of 11.3 compared with a full ACE simulation.

\end{abstract}


\maketitle

\section{Introduction}
\label{sec::introduction}
Molecular dynamics (MD) simulations have developed to a powerful tool to get insight into complex atomistic systems and to study their evolution in phase space.
The length and time scale, which can be covered, is strongly depended on the cost of energy- and force-computations between the particles.
The length and size of systems has been continuously extended over the years using parallel computing.
However, the accuracy of the interactions between particles in long time simulations has been limited in the past due to the approximative nature of classical force-field description.
In recent years the development and application of machine learning (ML) potentials (like the atomic-cluster expansion (ACE)\cite{ace}, moment tensor potentials\cite{moment_tensor_potentials}, Gaussian approximation potentials\cite{gaussian_approximation_potentials}, spectral neighbor analysis potentials\cite{spectral_neighbor_analysis_potentials} and neural network potentials\cite{neural_network_potentials}) have received large attention for the combination of computational performance and accuracy.
Accurate ML potentials are often based on large sets of density functional theory (DFT) data\cite{spectral_neighbor_analysis_potentials,seko_dft_reference,oord_dft_reference,pace}, used as a reference and therefore translate the accuracy of DFT calculations\cite{dft_overview} into dynamic simulations.
However, there is still a considerable performance gap between the evaluation of classical fields and ML potentials, e.g. the force calculation for a classical EAM potential and the ML-based ACE method differ by a factor of 100-1000\cite{pace}, which still limits the accessibility of time and length scales of systems, completely described by highly accurate descriptions and therefore pose a conflict between system size and precision.

Due to the high computational demands of accurate interaction models, there have been several attempts to couple low- to high-accurate
descriptions adaptively within in a simulation.
In QM/MM simulations quantum mechanical descriptions, based on DFT, are coupled to classical force fields\cite{qm_mm_coupling_hard_coded_zones1,Kermode_qmmm1,Kermode_qmmm2}.
Due to the large computational demand, the QM region is limited to a small sub-system.
On a coarser level, e.g. in soft matter\cite{soft_matter_polymer_adaptive_precision,AdReS_2008_force_based,soft_matter_polymer_adaptive_precision_3,Kermode_qmmm2}, an adaptive description has been introduced to reduce the number of degrees of freedom in a system to speed up the computations.
On even larger scales, where individual particles can be  combined to groups of particles in specific regions of the system coupling between classical atomistic simulations and continuum simulations (like material point method\cite{coupling_md_mpm_hard_coded_zones1} or finite element simulations\cite{coupling_md_fem_hard_coded_zones1} or with coarse-grained systems like in Refs. \cite{HAdReS_2013_energy_based,md_mpc_coupling_hard_coded_zones1}) are performed.

Due to the change of energy and force description in the two regions (high-/low-accuracy), a spatial zone is usually introduced, which
provides a smooth transition from one into the other region.
For consistency, either the energy or force is thereby weighted by a factor which changes from 0 to 1\cite{delle_site}.
Since the force is entering directly into the integration of motion, several approaches\cite{AdResS_2005_force_based,AdReS_2008_force_based,ml_ff_coupling_force_based} are based on a weighting of forces.
However, it was shown in Ref. \cite{delle_site} that this approach does not result in energy conservation, since the underlying potential cannot be properly reconstructed from the weighted forces.
An approach to generate energy conserving dynamics has been taken in Refs. \cite{energy_based_switching_qmmm,HAdReS_2013_energy_based,HAdReS_2015_energy_based}, where the weighting has been included into the Hamiltonian description.

With the advent of ML potentials, coupling of different interaction models shows strong potential for combining a high accurate description in small-to-medium sized subsystems with acceptable computational costs and a low accurate description in large subsystems, opening the path to long time and large length scale simulations including high accuracy, where it is needed.
Such a coupling between a ML-potential and a classical force-field description has been proposed in Ref. \cite{ml_ff_coupling_force_based} using a force-mixing approach to simulate a grand-canonical ensemble in the thermodynamic limit.

In contrast, in the present article we use an energy-mixing approach which is consistent with a Hamiltonian description and allows, in principle, the simulation of a microcanonical ensemble.
In so doing we introduce a method to couple a precise ML-potential for subregions of atoms of interest and a fast traditional potential for the remaining system components to overcome the conflict between system size and precision for classical atomistic simulations.
The atoms of interest are thereby automatically detected by a customizable detection mechanism and therefore the method works autonomously and self adaptively in space and time.
To further reduce the execution time we implemented the method into the parallel simulation engine LAMMPS\cite{lammps}.

The present paper is organized as follows: We first introduce our adaptive-precision model in \cref{sec::method}, whereas we present the energy-model in \cref{sec::energy_model}, discuss the group of interest detection in \cref{sec::switching_function} and the integration of motion in \cref{sec::integration_of_motion}.
With adaptive-precision and dynamic precision-selection in a parallel simulation comes the need for dynamic load balancing as the compute time changes over magnitudes between atoms.
Therefore, we present our dynamic load-balancing method in \cref{sec::load_balancing}.
We applied the introduced adaptive-precision mechanism and combined an EAM potential with an ACE potential for copper.
The input potentials are presented in \cref{sec::eam_ace} and the EAM potential is improved to be used in combination with the ACE potential.
Finally, we demonstrate in \cref{sec::demonstration} the capabilities in terms of precision and efficiency of the adaptive-precision copper potential for a nanoindentation of $4\times10^6$ atoms calculated for $\SI{100}{\pico\second}$ with LAMMPS\cite{lammps}.
In \cref{sec::conclusion} we conclude.

\section{Method}
\label{sec::method}
\subsection{Representation of energy and forces}
\label{sec::energy_model}
\begin{figure}[tb]
\begin{center}
\includegraphics[width=3.37in]{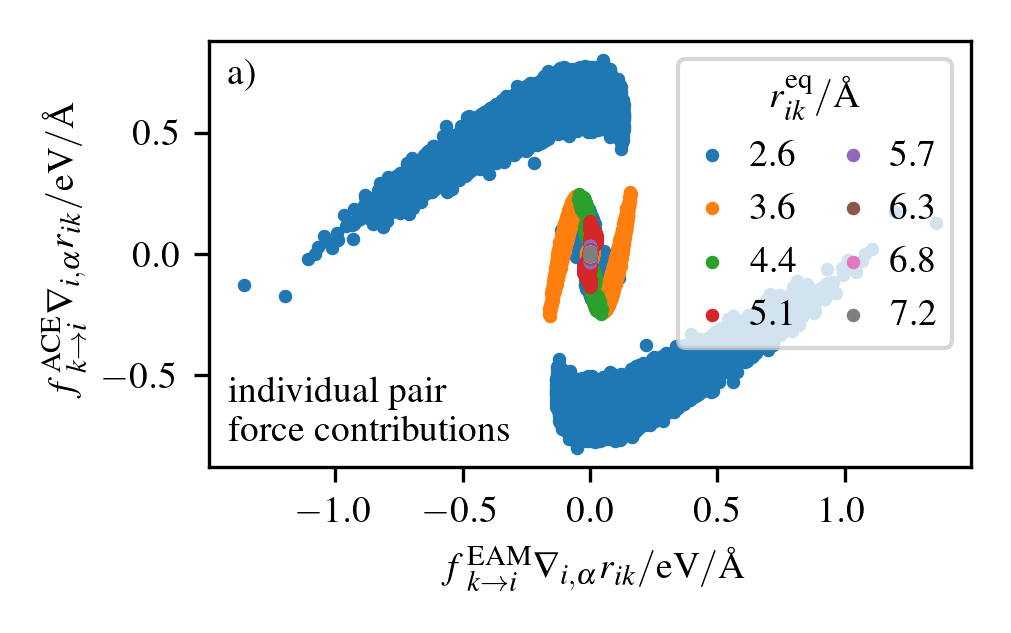}
\includegraphics[width=2.11in]{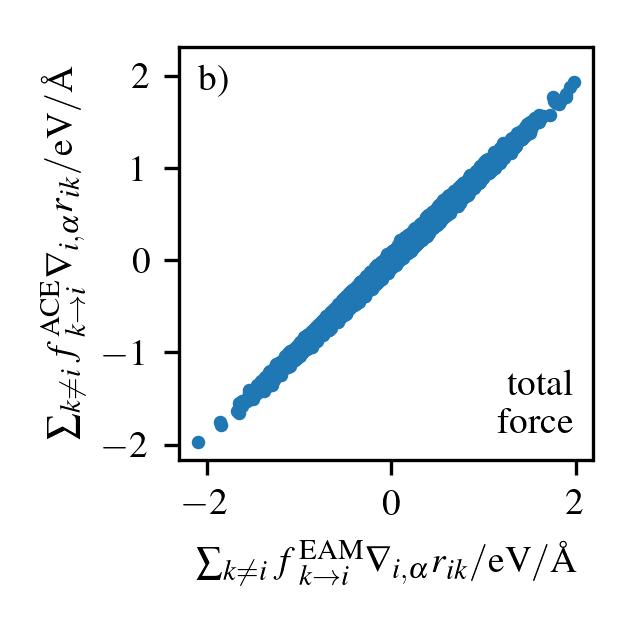}
\end{center}
\caption{\label{Fig::copper_f_ij_comparison}
a) Comparison of the individual pair force contributions $f_{k\rightarrow i}^{(M)}$ for a system of 13500 copper atoms at \SI{300}{\kelvin} with periodic-boundary conditions in all spatial dimensions $\alpha$.
The force contributions are colored according to the equilibrium distance $r_{ik}^\text{eq}$ of the corresponding atom pair $ik$.
The compared potentials are a modified EAM2 of Ref. \protect{\cite{mishin}} and ACE of Ref. \protect{\cite{pace}}, both of which are described in \cref{sec::eam_ace}.
b) Total force $F_i$ on an atom $i$ which is given by the sum over the individual pair force contributions of all neighbors $k$.
}
\end{figure}

The total energy of the system is given as $H = \sum_i (T_i + E_i) + E_\text{ext}$, where $T_i$ is the kinetic and $E_i$ the potential energy of atom $i$ and $E_\text{ext}$ are possibly existing external fields.
The energy of our adaptive-precision approach combines a precise and a fast energy, $E_i^\text{(p)}$ and $E_i^\text{(f)}$, per atom $i$.
We will use the atomic cluster expansion (ACE)\cite{ace} as precise and the embedded atom model (EAM)\cite{eam} as fast interatomic potential.
Precise and fast energies are combined with a continuous switching parameter $\lambda_i\in[0,1]$ per atom $i$, namely
\begin{equation}
E_i = \lambda_i E_i^\text{(f)} + (1-\lambda_i) E_i^\text{(p)}\,.
\label{eq::energy_hyb}
\end{equation}
During the course of a simulation the energy will be switched automatically from fast to precise or vice versa as required by the local atomic environment.
The switching parameter $\lambda_i$ needs to change continuously to prevent shocks by instant energy changes.
One can save computation time by using the combined energy $E_i$ compared to the precise energy $E_i^\text{(p)}$ since the precise calculation is only required for a subset of atoms.
One needs to make sure that the energies $E_i^\text{(s)}$ and $E_i^\text{(p)}$ are as similar as possible for the atoms, which are calculated with the fast potential, to prevent a systematic energy change due to the change of the switching parameter.
The force $F_i=-\nabla_i \sum_kE_k$ according to the energy model (\cref{eq::energy_hyb}) is
\begin{equation}
\begin{split}
F_i = \sum_k \Big(&- \lambda_k (\nabla_i E_k^\text{(f)}) - (1 - \lambda_k) (\nabla_i E_k^\text{(p)})\\
                  &+(\nabla_i\lambda_k) (E_k^\text{(p)} - E_k^\text{(f)})\Big)\label{eq::force_hyb}\,.
\end{split}
\end{equation}
It is important to note that $\lambda_i=0$ is not the only requirement for a precise force on atom $i$ since the force $F_i$ depends according to \cref{eq::force_hyb} on the switching parameters $\lambda_k$ of all atoms $k$ within the force cutoffs of the interatomic potentials.
For visualization, we define the force $f_{k\rightarrow i}^{(M)}$ from atom $k$ on atom $i$ according to the interatomic model $M$ as
$F^{(M)}_i = \sum_{k\neq i} f_{k\rightarrow i}^{(M)} \nabla_i r_{ik}$.
In contrast to the total force, the individual pair force contributions $f_{k\rightarrow i}^{(M)}$ of two different interatomic potentials are not necessarily correlated as shown in \cref{Fig::copper_f_ij_comparison}.

The adaptive-precision potential can be conservative and depend only on atomic distances $r_{ij}$ when a differentiable switching function $\lambda_i$ is used.
However, for a consistent force evaluation, gradients of $\lambda_k$ with respect to atomic positions need to be calculated (cmp. \cref{eq::force_hyb}).

\subsection{Adiabatic switching}
\label{sec::switching_function}
The switching function $\lambda$ changes the energy-precision (cmp. \cref{eq::energy_hyb}) and thereby also the force-precision during simulation dependent on the atomic environment.
This change of precision is ideally quasi-adiabatic to prevent perturbations of the system.
The switching function itself has generally no physical meaning, it is only an auxiliary function to determine the precision.
However, the force on atom $i$ depends according to \cref{eq::force_hyb} on $(\nabla_i\lambda_k)$ and thus not only on the value of the switching parameter but also on the calculation mechanism.
Therefore, an adiabatically slow change of the switching parameter is important to minimize the unphysical force contribution of $(\nabla_i\lambda_k)$.

The switching function needs a detection mechanism for particles that require a precise calculation.
This detection mechanism strongly depends on the simulation.
The centro-symmetry parameter\cite{csp} (CSP) detects e.g. defects and surface atoms whereas the common neighbor analysis (CNA)\cite{cna} can characterize grain boundary structures\cite{homer2019high}.
The common neighborhood parameter(CNP)\cite{tsuzuki2007structural} combines CNA and CSP.
CSP, CNA and CNP are implemented in LAMMPS\cite{lammps_csp,lammps_cna,lammps_cnp} and therefore accessible for a wider community.
For a coating like in Ref. \cite{luu2022nanoindentation} one can include only atoms of the same element like the central atom $i$ in the calculation of $\text{CSP}_i$ to make the interface look like a surface.

The switching parameters $\{\lambda_k\}$ of all atoms $k$ within the force cutoff of atom $i$ determine the force-precision of atom $i$ as discussed in \cref{sec::energy_model}.
Therefore, when an atom $i$ is detected for a precise calculation, the switching function needs to change also the switching parameters $\lambda_k$ of the neighboring atoms $k$ to ensure a precise force calculation of atom $i$.

To detect defects and surfaces, we assume here the CSP as detection mechanism.
However, it can be changed to other criteria without big implementation overhead.
The CSP is given as
\begin{equation}
\text{CSP}_{i}(t) = \sum_{j=1}^{N/2} \left( \vec{r}_{i,j}(t) + \vec{r}_{i,j+N/2}(t) \right)^2\,,
\label{eq::csp_1_i}
\end{equation}
whereas $N$ is the number of nearest neighbors and $j$ and $j+N/2$ correspond to a pair of opposite nearest neighbors of the central atom $i$.
How to identify these pairs is discussed in detail in \cref{sec::csp}.

The construction of our switching function, taking into account the requirements discussed, is described in \cref{sec::construction_switching_function}.

\subsection{Integration of motion}
\label{sec::integration_of_motion}
\subsubsection{Velocity-Verlet integration}
\label{sec:velocity_verlet}
With a known velocity $v_i(t)$, position $x_i(t)$, acceleration $a_i(t)$ and switching parameter $\lambda_i(t)$ of the last timestep $t$, one can propagate an atom in time.
A timestep of $\Delta t$ according to the Velocity-Verlet integration\cite{velocity_verlet} updated for the use with an adaptive-precision potential is given as,
\begin{subequations}
\label{eq::vv}
\begin{align}
&\text{1.}   \quad v_i\left(t + \frac{\Delta t}{2}\right) = v_i(t) + \frac{a_i(t)}{2} \Delta t\label{eq::vv_v1}\\
&\phantom{1.}\quad x_i\left(t + \Delta t\right) = x_i(t) + v_i\left(t + \frac{\Delta t}{2}\right) \Delta t\label{eq:vv_x}\\
&\text{2.}   \quad \lambda_i (t + \Delta t) = \lambda_i(\lambda_i(t),\{x(t+\Delta t)\})\label{eq::vv_lambda}\\
&\text{3.}   \quad a_i(t + \Delta t) = \frac{F_i\left(t + \Delta t\right)}{m_i}\label{eq::vv_a}\\
&\phantom{3.}\quad E_i(t + \Delta t) = E_i\left(t + \Delta t\right)\label{eq::vv_E}\\
&\text{4.}\quad v_i(t + \Delta t) = v_i\left(t + \frac{\Delta t}{2}\right) + \frac{1}{2} a_i(t + \Delta t) \Delta t\,,\label{eq::vv_v2}
\end{align}
\end{subequations}
whereas step 2 in \cref{eq::vv} is new and required due to the adaptive-precision potential.
The first velocity and position calculations for $v_i(t+\Delta t/2)$ and $x_i(t+\Delta t)$ use the given $\lambda_i(t)$.
A new lambda $\lambda_i(t+\Delta t)$ has to be calculated in the second step of the integration since the positions of the particles have changed and thus also their switching parameter.
Therefore, the afterwards calculated quantities $a_i(t+\Delta t)$, $E_i(t+\Delta t)$ and $v_i(t+\Delta t)$ depend on the new calculated lambda $\lambda_i(t+\Delta t)$.

\subsubsection{Avoiding unphysical force-contributions by $(\nabla_i\lambda_k)$}
\label{sec:local_thermostat}
The acceleration calculation according to \cref{eq::vv_a} requires according to \cref{eq::force_hyb} the knowledge of $(\nabla_i\lambda_k)$.
The force contribution through this gradient of the switching function should be small as a slowly changing switching function is used.
However, as discussed in \cref{sec::switching_function}, the switching function has generally no physical meaning, but only influences the required precision.
Furthermore, the need to change the switching parameter adiabatically slowly implies the usage of time averages whose gradient cannot be calculated consistently.
Thus, we want to neglect $(\nabla_i\lambda_k)$ in \cref{eq::force_hyb}, but this would violate energy-conservation.
The adaptive-precision potential is conservative for a constant set of $\{\lambda_i\}$ as the calculation of $\nabla_i\lambda_k$ is not required in this case.
This conservative reference system offers the possibility of neglecting the unphysical force contribution of $(\nabla_i\lambda_k)$ in an energy and momentum-conserving way.
We want to stress that if $\lambda$ is kept constant per atom over time we have a conservative dynamics and therefore conserve energy and momentum naturally.
Thus, we can use the system with constant switching parameters as reference system for each atom to apply a momentum-conserving correction for energy conservation which is described in the following.
Note that the presented correction approach is independent of the calculation of $\lambda$ and can be used for any switching function.

The energy difference $\Delta E_i$ of atom $i$ at the end of a time step between the conservative reference system and the system with updated switching parameters is caused by the change of the switching parameter.
To calculate this energy error $\Delta E_i$, one has to finish the calculation of an integration step for both sets of switching parameters.
As the forces of the reference system and the dynamic-$\lambda$ system depend on the same atomic positions, one can easily calculate both forces during the force calculation by just summing up two forces weighted with the corresponding switching parameter.
This energy difference affects both the kinetic energy change $\Delta T_i$ and the potential energy change $\Delta V_i$, namely
\begin{equation}
\Delta E_i = \Delta T_i + \Delta V_i\,.
\label{eq::Delta_E_i}
\end{equation}
The potential energy change $\Delta V_i$ affects only atoms whose switching parameter has changed (cmp. \cref{eq::energy_hyb}).
The kinetic energy change $\Delta T_i$ affects atoms in whose force cutoff a switching parameter has changed (cmp. \cref{eq::force_hyb}).
The calculation of $\Delta V_i$ and $\Delta T_i$ is described in \cref{sec::appendix_local_thermostat_error_calculation}.

We want to change the energy of particles in order to correct the energy error $\Delta E_i$ introduced by a changed $\lambda$ and thereby conserve the energy.
Since one cannot simply change the potential energy in a predictable way, we change the kinetic energy of particles.
The energy errors are calculated per particle $i$ and should also be corrected locally.
Therefore, we need a local thermostat which changes the kinetic energy of atoms by the measured energy error $\Delta E_i$.
The local thermostat rescales the momenta of a group of atoms $\Omega_i$ relative to the center-of-mass velocity by the factor $\beta(\Omega_i, \Delta E_i)$.
The rescaling conserves the momentum as the rescaling is relative to the centre-of-mass velocity.
The rescaling including the calculation of the rescaling factor $\beta(\Omega_i, \Delta E_i)$ is described in \cref{sec::appendix_local_thermostat_error_correction} in detail.
The integrator changes the velocity of particles of mass $m$ according to $\Delta v = \Delta t F / m$ due to a force $F$.
One can understand the energy-error correction as additional force which is just used at the end of a timestep, but not in the next integrator step.
The application of this local thermostat contains a stochastic component as direction of the effectively applied force is in direction of the relative momentum and thus stochastic.
Note that this local thermostat is not considered to provide a NVT ensemble.
It can be considered as an energy correction, but not primarily as temperature correction, i.e. the objective function is not the temperature, but the energy.
Thus, non-equilibrium simulations, non-energy conserving simulations like an NVT ensemble and simulations with external forces can be performed with the local thermostat.

\subsection{Load balancing}
\label{sec::load_balancing}
Time integrating the equations of motion for a set of atoms described by a short-range interatomic potential requires per atom only information from neighboring atoms within a cutoff distance.
Thus, LAMMPS uses distributed-memory parallelism via MPI\cite{lammps}.
The simulation box is divided into disjoint domains which cover the whole simulation box according to a spatial domain-decomposition approach\cite{spatial_domain_decomposition_lammps}.
Each MPI task, in the following denoted as processor, is assigned to one domain and administers all particles located in this domain.
Our adaptive-precision potential requires a highly different compute time per atom dependent on the used potential and that causes load-imbalances between the domains.
Hence, we present a load-balancing approach to dynamically change the domain sizes during a simulation.

We calculate the centro-symmetry parameter $\text{CSP}_{i}(t)$ according to \cref{eq::csp_1_i} and the switching parameter $\lambda$ according to \cref{eq::lambda_2_i} within the force-calculation routine to include both into the load balancing.
Thus, we perform four independent calculations during the force-calculation routine.
We execute firstly the calculation of the fast potential (FP), secondly the precise potential (PP), thirdly $\text{CSP}_{i}$ (CSP) and fourthly $\lambda_i$($\lambda$).
Since all four calculations are required only for a subset of particles as discussed in \cref{sec::atom_subgroups_load_balancing} in detail, the code inherently needs load balancing.
Required adjustments of the interatomic-potentials to allow effective load-balancing for adaptive-precision potentials are discussed in detail in \cref{sec::adjustments_for_load_balancing}.

We assume per processor $p$ and subroutine $\mathcal{X}$ an average work per particle
\begin{equation}
\rho_p^{(\mathcal{X})} = \tau_p^{(\mathcal{X})} / N_p^{(\mathcal{X})}\,,
\label{eq::rho_p_X}
\end{equation}
whereas the total duration $\tau_p^{(\mathcal{X})}$ and the number of calculations $N_p^{(\mathcal{X})}$ is measured within the force-calculation subroutine $\mathcal{X}$.
We rescale the work $\rho_p^{(\mathcal{X})}$ with a constant factor per processor $p$ in order to match the total work of our model with the also measured total calculation time $\tau_p$ of the processor.
Thereby, one can assign an individual load $\rho_i^\text{atom}$ to atom $i$ with possible contributions of $\rho_{p(i)}^\text{FP}$, $\rho_{p(i)}^\text{PP}$, $\rho_{p(i)}^\text{CSP}$ and $\rho_{p(i)}^\lambda$.
The load $\rho_i^\text{atom}$ is used as input for a load-balancing algorithm.
We use a staggered grid\cite{rene_masterarbeit} which is used for particle simulations\cite{rene_staggered_conference_paper} and which is implemented in the load-balancing library ALL\cite{ALL}.

\section{Adaptive-precision potential}
\label{sec::eam_ace}
\subsection{Atomic cluster expansion}
\label{sec::ace}
We use the atomic cluster expansion (ACE) of Ref. \cite{pace} as precise potential as it is a modern ML-potential with a good representation of its DFT reference data.
The energy $E_i$ of an atom $i$ described by the atomic cluster expansion\cite{ace} is
\begin{equation}
E_i^\text{ACE} = \phi_i^{(1)} + \sqrt{\phi_i^{(2)}}\,,
\end{equation}
whereas the functions $\phi_i^{(p)}$ are expanded as
\begin{equation}
\phi_i^{(p)} = \sum_{\bm{v}} c_{\bm{v}}^{(p)} B_{i\bm{v}}\,,
\label{eq::energy_ace}
\end{equation}
where $B_{i\bm{v}}$ are product basis functions which describe the atomic environment and $c_{\bm{v}}^{(p)}$ are fitted expansion coefficients with the multi-indices $\bm{v}$.
We use the copper ACE potential from Ref. \cite{pace}.

\subsection{Embedded atom model}
\label{sec::eam}
\begin{figure}[]
\begin{center}
\includegraphics[width=3.37in]{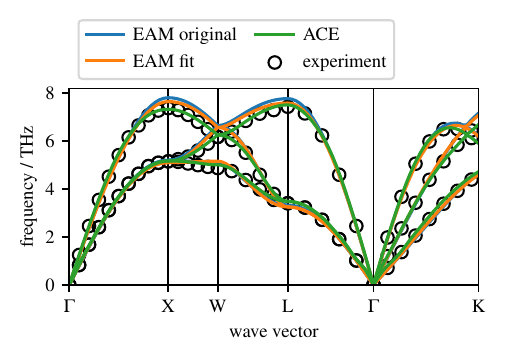}
\end{center}
\caption{\label{Fig::copper_fit_eam_phonon_spectrum}
Phonon spectra calculated with ASE\protect{\cite{atomic_simulation_enviroment}} of the fitted EAM potential, original EAM potential, the ACE potential and experimental values\protect{\cite{experimental_phonon_spectrum_copper}} measured by neutron diffraction at the temperature $\SI{80}{\kelvin}$.
}
\end{figure}

We use the EAM2 potential of Ref. \cite{mishin} as starting point for the fast potential since it is widely used\cite{mishin_use_1,mishin_use_2,mishin_use_3,mishin_use_4}.
The energy of an atom $i$ described with the embedded atom model\cite{eam} (EAM) is
\begin{equation}
E_i^\text{EAM} = \xi\left(\sum_{j\neq i}\zeta(r_{ij})\right) + \frac{1}{2} \sum_{j\neq i} \Phi(r_{ij})\,,
\label{eq::energy_eam}
\end{equation}
where $\xi$ is the embedding function, $\zeta$ the electron density and $\Phi$ a pair potential.
We use a copper potential from Ref. \cite{mishin} which is described in \cref{sec::Mishins_EAM_potential}.
We need to optimize the EAM potential to minimize systematic energy differences between the used EAM and ACE potentials.
Furthermore, the optimization gives the possibility to improve the forces as well.
The EAM potential is fitted for applications with surfaces, vacancies and interstitials which are detected for a precise calculation and therefore do not need to be described by the fast and less accurate EAM potential.
To correct an energy offset, we introduce $\xi_0^\text{Fit}$ to the original embedding function $\xi(x)$ (\cref{eq::mishin_embedding_function}), namely $\xi^\text{Fit}(x) = \xi(x) + \xi_0^\text{Fit}$ and use $\xi^\text{Fit}(x)$ instead of $\xi(x)$.
The set of parameters $\mathcal{A}$ (cmp. \cref{sec::Mishins_EAM_potential}) is optimized by minimizing a loss function $\mathcal{L}$ with atomicrex\cite{atomicrex}.
The loss function includes atomic energies $E_{s,i}$ and forces $F_{s,i}$ of different structures $s$ with $N_s$ atoms as well as the scalar properties $A_o$.
We use the lattice parameter $a_0^\text{FCC}$, the cohesive energy and elastic constants as scalar properties.
The target values denoted with 'targ' are the corresponding values calculated with the highly accurate ACE potential.
The predicted values denoted with 'pred' are calculated with the current set of parameters $\mathcal{A}$.
The differences between predicted and target values are per quantity weighted with a tolerance $\delta^\text{tol}$.
The used target values and tolerances of the loss function are listed in \cref{tab::target_values_copper}.
\begin{table}
\caption{\label{tab::target_values_copper}Target values with tolerance used in the loss function \cref{eq::atomicrex_loss_function} for the optimization of the EAM potential.}
\centering
\begin{tabular}{lrr}
\hline\hline
bulk property                      & target value                                               & tolerance $\delta^\text{tol}$           \\\hline
lattice parameter $a_0^\text{FCC}$ & \round{4}{3.63090872983321}\,\si{\angstrom}                & \SI{0.001}{\angstrom}                   \\
cohesive energy $E_\text{coh}$     & \round{4}{-3.69947432079432}\,\si{\electronvolt\per atom}  & \SI{0.01 }{\electronvolt\per atom}      \\
bulk modulus $B$                   & \round{1}{138.240089803346}\,\si{\giga\pascal}             & \SI{1    }{\giga\pascal}                \\
elastic constant $C_{11}$          & \round{1}{173.7688102492}\,\si{\giga\pascal}               & \SI{1    }{\giga\pascal}                \\
elastic constant $C_{12}$          & \round{1}{120.4757295789}\,\si{\giga\pascal}               & \SI{1    }{\giga\pascal}                \\
elastic constant $C_{44}$          & \round{1}{77.78014061342}\,\si{\giga\pascal}               & \SI{1    }{\giga\pascal}                \\
force $F_\text{md}$                & MD simulation                                              & \SI{0.01 }{\electronvolt\per\angstrom}  \\
atomic energy $E_\text{md}$        & MD simulation                                              & \SI{0.01 }{\electronvolt}               \\
\hline\hline
\end{tabular}
\end{table}
The minimized loss function is
\begin{equation}
\begin{split}
\mathcal{L}(\mathcal{A}) =& \sum_{s\in\mathcal{S}} \frac{1}{N_s} \Bigg(\sum_{i=1}^{N_s} \left(\frac{E_{s,i}^\text{targ} - E_{s,i}^\text{pred}}{\delta^\text{tol}_{E_\text{md}}}\right)^2\\
&\phantom{\sum_{s\in\mathcal{S}} \frac{1}{N_s} \Bigg(} + \sum_{i=1}^{N_s} \left(\frac{\lVert F_{s,i}^\text{targ} - F_{s,i}^\text{pred}i\rVert}{\delta^\text{tol}_{F_\text{md}}}\right)^2\Bigg)\\
& +\sum_{o\in\mathcal{O}} \left(\frac{A^\text{targ}_o - A^\text{pred}_o}{\delta^\text{tol}_o} \right)^2\,,
\label{eq::atomicrex_loss_function}
\end{split}
\end{equation}
with the set of scalar observables $\mathcal{O}$ and the set of structures $\mathcal{S}$.
We use six snapshots of a NVE simulation of copper with $10\times10\times10$ unit cells at \SI{300}{\kelvin} and periodic boundary conditions as structures for the fit.
These structures are sufficient since we start with an already tested and validated EAM potential and the optimized EAM potential is for use only with atomic configurations similar to the used structures.
We start the fitting with the parameters of EAM2 from Ref. \cite{mishin} and $\xi_0^\text{Fit}=\SI{0}{\electronvolt}$.
In a first run, only the energy offset $\xi_0^\text{Fit}$ is fitted.
Afterwards all parameters including $\xi_0^\text{Fit}$ are fitted.
The optimized parameters are listed in \cref{sec::Mishins_EAM_potential} in \cref{tab::copper_fit_300K_Mishin_parameters}.
Scalar properties calculated with the optimized EAM potential are shown in \cref{tab::properties_eam_copper_fit}.
\begin{table*}
\caption{
\label{tab::properties_eam_copper_fit}Comparison of quantities calculated with EAM, optimized EAM and ACE with DFT and experimental reference values.
}
\centering
\begin{tabular}{lrrrrr}
\hline\hline
                                                           & ACE                          & EAM                          & EAM                          & DFT  & Exp.    \\
                                                           &                              & original                     & fit                          &      &         \\\hline
\normalsize{Lattice Constant / \si{\angstrom}}             & \round{4}{3.63090872983321 } & \round{4}{3.61492887282983 } & \round{4}{3.63045256622797 } &      &         \\
\normalsize{Cohesive Energy / \si{\electronvolt}}          & \round{4}{-3.69947432079432} & \round{4}{-3.55786833489833} & \round{4}{-3.69569662929878} &      &         \\
\normalsize{Vacancy Formation Energy / \si{\electronvolt}} & \round{4}{1.12853673484551 } & \round{4}{1.27349604826122 } & \round{4}{1.0588710171005  } & 1.07\cite{pace} & 1.27 \cite{copper_exp_vacancy_formation}   \\
\normalsize{Surface Energy 111 / \si{\joule\metre^{-2}}}   & \round{4}{1.35660821328453 } & \round{4}{1.24430605633688 } & \round{4}{0.88024354655695 } & 1.36\cite{pace} &         \\
\normalsize{Surface Energy 100 / \si{\joule\metre^{-2}}}   & \round{4}{1.51112761487184 } & \round{4}{1.35153002045502 } & \round{4}{0.97649687173991 } & 1.51\cite{pace} &         \\
\normalsize{Surface Energy 110 / \si{\joule\metre^{-2}}}   & \round{4}{1.58893923614342 } & \round{4}{1.48217604297006 } & \round{4}{1.12251374356068 } & 1.57\cite{pace} &         \\
Interstitial Formation Energy \\
\normalsize{\quad 100-dumbbell / \si{\electronvolt}}       & \round{4}{3.1024697749345  } & \round{4}{3.08094456110825 } & \round{4}{2.96618403515875 } & 3.10\cite{copper_dft_point_defects} & 2.8-4.2\cite{copper_exp_interstitial_db} \\
\normalsize{\quad octahedral / \si{\electronvolt}}         & \round{4}{3.3392327783941  } & \round{4}{3.2558783098782  } & \round{4}{3.13034901463887 } & 3.35\cite{copper_dft_point_defects} &         \\
\normalsize{\quad tetrahedral / \si{\electronvolt}}        & \round{4}{3.62169967073448 } & \round{4}{3.5655217348783  } & \round{4}{3.42395856517813 } & 3.64\cite{copper_dft_point_defects} &         \\
\normalsize{Elastic Constant C11  / \si{\giga\pascal}}     & \round{1}{173.768810251394 } & \round{1}{171.453901566772 } & \round{1}{171.288636692165 } &  177\cite{pace} & 177 \cite{copper_exp_elastic_constants}    \\
\normalsize{Elastic Constant C12  / \si{\giga\pascal}}     & \round{1}{120.475729579322 } & \round{1}{124.157217644317 } & \round{1}{120.145974948912 } &  132\cite{pace} & 125 \cite{copper_exp_elastic_constants}    \\
\normalsize{Elastic Constant C44  / \si{\giga\pascal}}     & \round{1}{77.7801406113201 } & \round{1}{76.2101973267168 } & \round{1}{80.5610917066166 } &   82\cite{pace} &  81 \cite{copper_exp_elastic_constants}    \\
\hline\hline
\end{tabular}
\end{table*}
Lattice constant and cohesive energy of the optimized EAM potential match through the optimization.
The elastic constants of ACE are in good agreement with both EAM and optimized EAM potential.
The phonon spectra are compared in \cref{Fig::copper_fit_eam_phonon_spectrum} and in are good agreement.

\subsection{Combined EAM and ACE potential}
\label{sec::combined_eam_ace_potential}
We combine the ACE potential and the optimized EAM potential to an adaptive-precision copper potential according to \cref{eq::energy_hyb} with the switching function $\lambda$, which is described in \cref{sec::switching_function,sec::construction_switching_function}.
There are parameters in three components, firstly in the mechanism used to detect atoms that require a precise calculation, secondly in the switching function and thirdly in the local thermostat used to avoid unphysical force contributions by $(\nabla_i\lambda_k$).
We discuss how to select these parameters in \cref{sec::parameter_selection}, whereas the parameter set is listed in \cref{tab::model_parameters}.
We refer to the adaptive-precision potential with this parameter set as Hyb1.

\section{Demonstration}
\label{sec::demonstration}
We calculated a nanoindentation with 4 million atoms and a (100)-surface at $\SI{300}{\kelvin}$ for $\SI{100}{\pico\second}$ on JURECA-DC\cite{jureca} with the adaptive-precision copper potential Hyb1 (cmp. \cref{tab::model_parameters}) to analyze precision and saved computation time compared to a full ACE simulation.

\subsection{Precision}
\label{sec::precision_results}
\begin{figure}[tb]
\begin{center}
\includegraphics[width=3.37in]{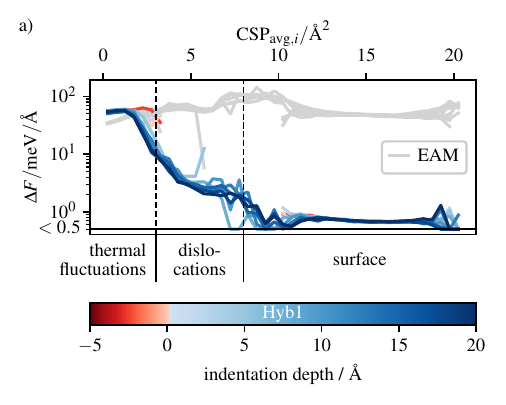}
\includegraphics[width=3.37in]{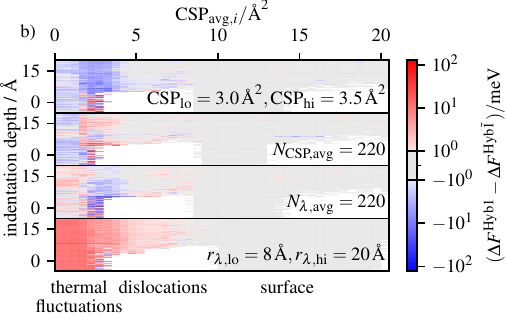}
\end{center}
\caption{\label{Fig::force_precision_simulations}
a) Force error $\Delta F^\text{Hyb1} = \text{RMSE}(F_{i{\alpha}}^\text{Hyb1}-F_{i\alpha}^\text{ACE})$ of rescaled atoms dependent on $\text{CSP}_{\text{avg},i}$ and indentation depth of a nanoindentation.
b) Force error $\Delta F^\text{Hyb1}$ compared with the force error $\Delta F^{\text{Hyb}\tilde{1}}$ of another simulation with the parameter set Hyb$\tilde{1}$, whereas the parameters in the lower right corner of the corresponding plot are different than in Hyb1.
The force error difference $\Delta F^\text{Hyb1}-\Delta F^{\text{Hyb}\tilde{1}}$ is color-coded, whereas blue and red corresponds to a smaller force error of Hyb1 and Hyb$\tilde{1}$ respectively.
The simulations for a) and the lowest plot of b) are calculated in a cubic system with $100^3$ unit cells, $68^3$ unit cells are used for the remaining simulations.
}
\end{figure}
The force difference $\Delta F^\text{Hyb1} = \text{RMSE}(F_{i\alpha}^\text{Hyb1}-F_{i\alpha}^\text{ACE})$ of rescaled atoms, including the theoretical rescaling force according to \cref{eq::additional_force_contribution}, of the adaptive-precision potential Hyb1 compared to ACE forces during the nanoindentation is shown in \cref{Fig::force_precision_simulations}a.
The force difference $\Delta F^\text{Hyb1}$ is calculated from force components $F_{i\alpha}$ in the spatial dimension $\alpha$.
For defects and surface atoms, the forces are within a tolerance up to $\SI{10}{\milli\electronvolt/\angstrom}$ compared to the precise ACE forces for most cases while the exact potential energy of ACE is used according to \cref{eq::energy_hyb}.

We changed single parameters of the parameter set Hyb1 and performed simulations with otherwise identical parameters to perform a sensitivity analysis on results of the selected parameters.
The influence of the changed parameters is shown in \cref{Fig::force_precision_simulations}b.
Increasing the thresholds $\text{CSP}_\text{lo}$ and $\text{CSP}_\text{hi}$ by $\SI{0.5}{\angstrom^2}$ decreases the force precision for thermally fluctuating atoms and dislocations with $\text{CSP}_{\text{avg},i}<\SI{4.5}{\angstrom^2}$.
The decreased force precision for the dislocations is expected as these are the atoms which are simulated with the fast potential due to the increased CSP thresholds.
Thus, the only advantage of increasing the CSP thresholds is performance.
Doubling the averaging time of the centro-symmetry parameter $N_{\text{CSP},\text{avg}}$ decreases the force precision on thermally fluctuating atoms.
The force precision of dislocations increases in most snapshots.
As dislocations are in contrast to the surface not stationary, increasing the CSP averaging time may reduce the detection possibility for moving dislocations.
Thus, one can adjust $N_{\text{CSP},\text{avg}}$ dependent on the expected dislocation dynamics.
Doubling the averaging time of the switching parameter $N_{\lambda,\text{avg}}$ decreases the force precision of dislocations with $\text{CSP}_{\text{avg},i}<\SI{4}{\angstrom^2}$, but increases the force precision of dislocations with a larger time-averaged CSP.
The force precision for dislocations decreases with an increasing $\text{CSP}_{\text{avg},i}$ according to \cref{Fig::force_precision_simulations}a and this effect is enhanced by doubling $N_{\lambda,\text{avg}}$.
Thus, increasing $N_{\lambda,\text{avg}}$ is not beneficial.
Increasing the cutoffs $r_{\lambda,\text{lo}}$ and $r_{\lambda,\text{hi}}$ of $\lambda_{\text{min},i}$ by $\SI{4}{\angstrom}$ and $\SI{8}{\angstrom}$ respectively, increases the number of precise calculations and thus the force precision for all atoms apart from surface atoms which are already treated with the precise potential.
The only disadvantage is the increased compute time due to more precise calculations.
Therefore, the cutoffs of $\lambda_{\text{min},i}$ can be adjusted dependent on the performance requirements.

\subsection{Computational efficiency}
\label{sec::results_load_balancing}
The processor $p$ with the highest work $\tau_p^\text{max}$ restricts the speed of a simulation since faster processors need to wait due to communication and synchronization.
Thus, the imbalance, as defined in LAMMPS\cite{lammps_fix_balance},
\begin{equation}
I=\tau_p^\text{max}/\langle\tau_p\rangle_p
\label{eq::imbalance}
\end{equation}
of the force calculation time $\tau_p$ gives a measure for the quality of the load balancing, whereas an imbalance of 1 corresponds to a perfectly balanced system.
The mean imbalance in the nanoindentation is 1.41 with a standard deviation of 0.26.
Dynamic load-balancing of an adaptive-precision potential is challenging as the work distribution changes drastically when the precision of atoms is switched from fast to precise or vice versa as discussed in \cref{sec::dynamic_load_balancing_details} in more detail.
The average work per atom for the four subprocesses is $\langle\rho_{p,t}^\text{FP}\rangle_{p,t} = \SI{9.6}{\micro\second}$, $\langle\rho_{p,t}^\text{PP}\rangle_{p,t} = \SI{1.7}{\milli\second}$, $\langle\rho_{p,t}^\lambda\rangle_{p,t} = \SI{48.6}{\micro\second}$ and $\langle\rho_{p,t}^\text{CSP}\rangle_{p,t} = \SI{5.6}{\micro\second}$, whereas the processor-dependency is discussed in \cref{sec::processor_dependency_work}.

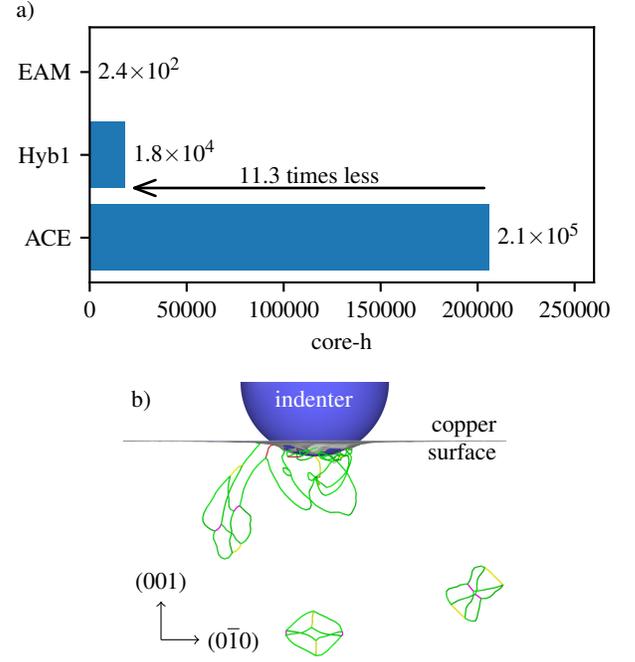
\begin{figure}
\includegraphics[width=3.37in]{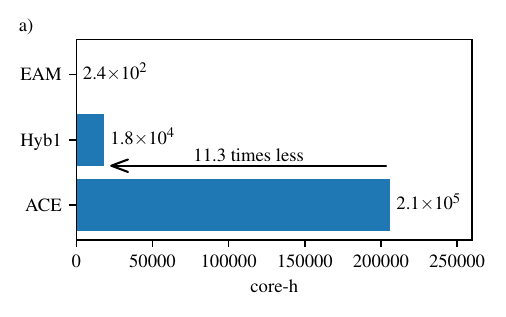}
\begin{tikzpicture}[x=1in,y=1in]
  \node[anchor=north west,inner sep=0] at (0,0) {
    \includegraphics[width=2.00in]{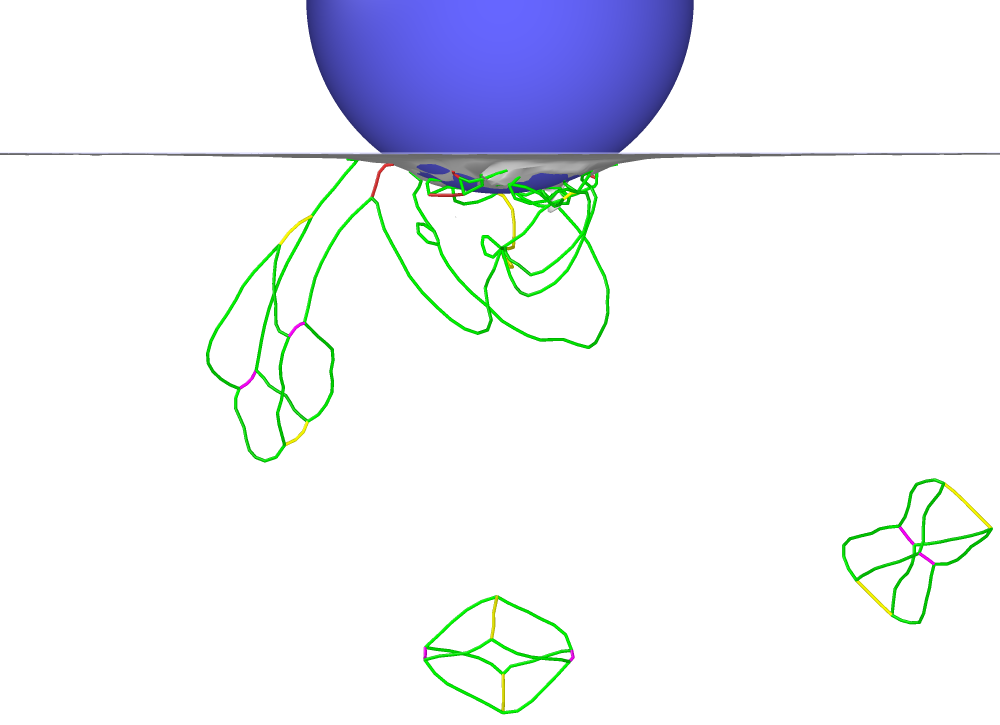}
  };
  \node [anchor = north west] at (0,0) {b)};
  \node [anchor = north] at (1,0) {\textcolor{white}{indenter}};
  \node [anchor = north east, align=right] at (2,-0.16) {copper\\surface};
  \node [coordinate] (tripod_origin) at (0.2,-1.35) [] {};
  \draw[->] (tripod_origin) -- ($(tripod_origin) + (0.2,0)$) node[anchor=west] {(0$\overline{1}$0)};
  \draw[->] (tripod_origin) -- ($(tripod_origin) + (0,0.2)$) node[anchor=south] {(001)};
\end{tikzpicture}
\caption{\label{Fig::speedup_nanoindentation}
a)
Total computation time of a nanoindentation with 4 million atoms ($100^3$ unit cells) simulated for $\SI{100}{\pico\second}$ with the adaptive-precision potential Hyb1 compared to ACE and EAM nanoindentations.
EAM, hybrid and ACE simulations are calculated on 128, 384 and 2048 cores of JURECA-DC\protect{\cite{jureca}} respectively.
b)
Visualization of the dislocation line defects which occur in a nanoindentation.
The visualization is done with OVITO\protect{\cite{ovito}} and the dislocations are identified by the dislocation analysis\protect{\cite{ovito_dxa}} of OVITO.
}
\end{figure}
The motivation to use an adaptive-precision potential is to save computation time compared to a precise potential while preserving accuracy in regions of the simulation where it is required.
The nanoindentation with 4 million atoms calculated for $\SI{100}{\pico\second}$ on JURECA-DC\cite{jureca} with the adaptive-precision potential Hyb1 (cmp. \cref{tab::model_parameters}) shows a speedup of 11.3 compared to a full ACE simulation like visualized in \cref{Fig::speedup_nanoindentation}a.
We calculated the surface and dislocations, which developed during the nanoindentation like visualized in \cref{Fig::speedup_nanoindentation}b, precisely with the ACE-potential and saved computation times for the remaining atoms by using the fast EAM-potential.
Note that the amount of saved computation time strongly depends on the number of not required evaluations of the precise and expensive potential as discussed in \cref{sec::speedup_equilibration} using the equilibration of the nanoindentation as example.

\section{Conclusion}
\label{sec::conclusion}
We introduced a method which allows to compute adaptive-precision interatomic potentials to speedup atomistic simulations with a region or atoms of special interest.
The atoms of special interest are automatically detected by a customizable detection mechanism and simulated with the precise potential while the fast potential is used for the remaining atoms.
The energy model (\cref{eq::energy_hyb}) ensures precise energies for the detected atoms while the switching function (cmp. \cref{sec::switching_function}) is used to get also precise forces.
We presented a load-balancing method with subroutine specific per-atom works per processor using a staggered grid to prevent the load-balancing issues arising by combining two potentials of different cost.
We demonstrated the presented adaptive-precision method by creating the copper potential Hyb1 (\cref{tab::model_parameters}) from an EAM and an ACE potential.
We explained how the fast EAM potential is optimized to match bulk properties with the precise ACE potential.

A nanoindentation with 4 million atoms calculated for $\SI{100}{\pico\second}$ was used to demonstrate the capabilities of the method.
The achieved accuracy of Hyb1 for the detected atoms of interest during the nanoindentation is for forces $\SI{10}{\milli\electronvolt/\angstrom}$ and for potential energies $\SI{0}{\milli\electronvolt}$.
The nanoindentation showed a speedup of 11.3 compared to a full ACE simulation.
Note that the amount of computation time that can be saved depends on the number of not required precise calculations and therefore on the simulation setup and the detection mechanism for atoms of interest.

The application of an adaptive-precision potential for other situations where the highest precision is required only locally but not globally like interfaces, cracks and grain boundaries is straightforward while the generalization of the detection mechanism for non-crystalline systems like amorphous solids is a natural next question.
Automated training of the fast potential would improve the usability of adaptive-precision potentials.
The present work is based on a CPU-version of our method.
A GPU-version is planned for the future.

\begin{acknowledgments}
We would like to thank Y. Lysogorskiy for help with the ML-PACE package in LAMMPS and for valuable comments on the integration of the switching parameter within the ACE calculation.
The authors gratefully acknowledge computing time on the supercomputer JURECA\cite{jureca} at Forschungszentrum Jülich under grant no. 28990 (hybridace).
\end{acknowledgments}

\section*{Author declarations}
\subsection*{Competing interests}
The authors declare no competing interests.

\subsection*{Author Contributions}
\textbf{David Immel:}
Formal analysis;
Investigation;
Methodology (equal);
Resources (equal);
Software;
Validation;
Visualization;
Writing - original draft;
\textbf{Ralf Drautz:}
Conceptualization (equal);
Methodology (equal);
Supervision (equal);
Writing - review \& editing (equal);
\textbf{Godehard Sutmann:}
Conceptualization (equal);
Methodology (equal);
Resources (equal);
Supervision (equal);
Writing - review \& editing (equal);

\section*{Data availability}
The data that support the findings of this study are available from D.I. upon reasonable request.
\bibliography{references.bib}

\appendix
\section{Centro-symmetry parameter}
\label{sec::csp}
To calculate the CSP according to Ref. \cite{csp}, initially reference vectors $\vec{r}_{i,j,\text{ref}}$ from the central atom $i$ to its closest neighboring atoms $j$ are determined in an undistorted lattice.
In the distorted lattice, one uses the distorted vectors $\vec{r}_{i,j}$ closest in distance to the undistorted reference vectors to calculate the centro-symmetry parameter
$\text{CSP}_i^\text{ref} = \sum_{j=1}^{N/2} (\vec{r}_{i,j} + \vec{r}_{i,j+N/2})^2$,
whereas $j$ and $j+N/2$ correspond to a pair of opposite nearest neighbors of the central atom $i$.
The set of distorted vectors used to calculate $\text{CSP}_i^\text{ref}$ may contain a neighbor multiple times and non-nearest neighbors.

\begin{figure}[b]
\begin{center}
\includegraphics[width=0.98in]{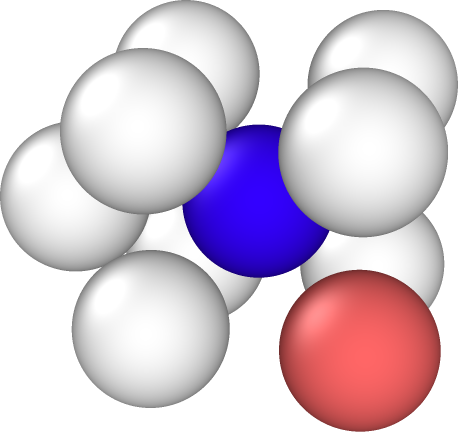}
\end{center}
\caption{\label{Fig::csp_calculation_problem_bcc_tungsten} Snapshot of BCC Tungsten at 300\,K. The eight nearest neighbors (white) of the central atom (blue) do not contain all neighboring atoms required to calculate the centro-symmetry parameter $\text{CSP}^\text{ref}$. The 9th-nearest neighbor(red) is missing. Thus, one needs to evaluate more than the 8 nearest neighbors for the $\text{CSP}^\text{lmp}$-calculation.}
\end{figure}
LAMMPS\cite{lammps_csp} uses a different definition for the centro-symmetry parameter $\text{CSP}^\text{lmp}$ which does not require a reference system.
At first, one searches the set $\mathcal{N}_i$ of the $N$ nearest neighboring atoms of the central atom $i$, whereas $N$ is the number of nearest neighbors in the undistorted lattice.
Afterwards, the $N/2$ pairs with the smallest $|\vec{r}_{i,j_1} + \vec{r}_{i,j_2}|$ with $j_1,j_2\in\mathcal{N}_i$ are used to calculate the centro-symmetry parameter.
This definition may also contain duplicates, but relevant non-nearest neighbors may be neglected.
This neglection of non-nearest neighbors is relevant for the computed $\text{CSP}^\text{lmp}$ of Tungsten in a BCC lattice at $\SI{300}{\kelvin}$ like shown in \cref{Fig::csp_calculation_problem_bcc_tungsten}.
The effect of this neglection is reduced by searching $N+N_\text{buffer}$ rather than $N$ nearest neighbors for $\mathcal{N}_i$.
Thus, we calculate $\text{CSP}^\text{lmp}$ according to the LAMMPS-definition, but use $N+N_\text{buffer}$ atoms for the identification of the $N/2$ relevant opposite-neighbor pairs denoted as ${j, j+N/2}$.

The switching parameter needs to detect atoms for the precise calculation precisely to prevent too many evaluations of the precise and expensive potential.
This requirement is fulfilled by the CSP since only neighboring atoms similar to the expected neighbors in an undistorted lattice are used for the calculation.
Missing or displaced atoms in the second neighbor-shell of the central atom do not influence the CSP.
Hence, the nearest neighbor-shell of a defect, vacancy or surface atom is detected, but the second-nearest neighbor-shell is not.
The latter behavior might not be useful when low angle symmetric tilt grain boundaries, for example, are of interest as only a subset of atoms may be detected\cite{coleman2013virtual}.
However, one can use the CSP to detect some grain boundaries\cite{li2019atomistic}.

\section{Construction of the switching function}
\label{sec::construction_switching_function}
The centro-symmetry parameter according to \cref{eq::csp_1_i} detects defects and surface atoms, which need to be calculated by the precise potential.
Therefore, we use the CSP as starting point to construct the switching function for a nanoindentation.
The centro-symmetry parameter fluctuates over time as atoms fluctuate dependent on their temperature.
Since the thermal fluctuations of atoms should not be detected, we introduce a moving average
\begin{equation}
\text{CSP}_{\text{avg},i}(t) = \frac{1}{N_{\text{CSP},\text{avg}}} \sum_{n=1}^{N_{\text{CSP},\text{avg}}} \text{CSP}_{i}(t - n \Delta t)
\label{eq::csp_2_i}
\end{equation}
of $N_{\text{CSP},\text{avg}}$ timesteps $\Delta t$.
The time-averaged centro-symmetry parameter can be used to calculate a switching parameter $\lambda_{0,i}$ as
\begin{equation}
\lambda_{0,i}(t) = 1 - f^\text{(cut)} \left(\frac{\text{CSP}_{\text{avg},i}(t) - \text{CSP}_\text{lo}}{\text{CSP}_\text{hi} - \text{CSP}_\text{lo}} \right)\,,
\label{eq::lambda_1_i}
\end{equation}
where we use the radial function
\begin{equation}
f^\text{(cut)}(x) =
\left\{
\begin{array}{ll}
1 & \text{for } 0 > x\,,\\
1 - 1.875x + 1.25x^3 - 0.375x^5 & \text{for }  0 \leq x \leq 1\,,\\
0& \text{for } x > 1  \,.
\end{array}
\right.
\end{equation}
$f^\text{(cut)}$ is taken from PACE\cite{pace} as the first and second derivative of $f^\text{(cut)}$ are smooth at 0 and 1.
$\lambda_{0,i}(t) = 1$ applies for all atoms with $\text{CSP}_{\text{avg},i}(t) \leq \text{CSP}_\text{lo}$.
The use of a continuous switching function for $\text{CSP}_{\text{avg},i}\in[\text{CSP}_\text{lo}, \text{CSP}_\text{hi}]$ ensures a smooth transition from the fast interatomic potential to the precise one and vice versa for all atoms.

One can change the argument of $f^{\text{(cut)}}$ in \cref{eq::lambda_1_i} to use a different detection mechanism and use the following definitions of the switching function as presented without further changes.
The argument must be 0 for atoms to be calculated exactly and 1 for atoms to be calculated quickly.

Since the force on an atom depends on all switching parameters within its force cutoff as discussed in \cref{sec::energy_model}, we introduce
\begin{equation}
\begin{split}
\lambda_{\text{min},i}(t) = \text{min}\Big(\Big\{&1 - (1 -\lambda_{0,j}(t)) f^\text{(cut)}\left(\frac{r_{ij}(t)-r_{\lambda,\text{lo}}}{r_{\lambda,\text{hi}} - r_{\lambda,\text{lo}}}\right)\\
&: j \in \Omega_{\lambda,i} \Big\}\Big)\,,
\label{eq::lambda_2_i}
\end{split}
\end{equation}
where $\Omega_{\lambda,i} = \{ j : r_{ij} < r_{\lambda,\text{hi}}\}$ is the set of neighboring atoms.
As $\lambda_{0,i}\in[0,1]$ applies, $\lambda_{\text{min},i}\in[0,1]$ follows.
For both calculation and understanding of $\lambda_{\text{min},i}(t)$ it is useful to start with $\lambda_{0,i}(t)$ as initial value for the minimum search.
Atoms $j$ that were not detected for a precise calculation have $\lambda_{0,j}(t)=1$.
Their contribution to the minimum search is $1$ independent of $r_{ij}$ and thus negligible.
Only neighbors $j$ that were detected for a precise calculation and thus have $\lambda_{0,j}(t)<1$ and $\lambda_{0,i}(t)$ can be the searched minimum.
Atoms with $\lambda_{0,i}(t)=0$ set $\lambda_{\text{min},j}(t)=0$ for all neighboring atoms $j$ with $r_{ij}\leq r_{\lambda,\text{lo}}$ and may set $\lambda_{\text{min},j}\in(0,1)$ for neighboring atoms with $r_{ij}\in(r_{\lambda,\text{lo}}, r_{\lambda,\text{hi}})$.
Thus, \cref{eq::lambda_2_i} decreases the switching parameter for neighbors of atoms, which need to be calculated precisely, to ensure a higher fraction of precise force contributions for these atoms.
Since the radial function $f^\text{(cut)}(r_{ij})$ used in \cref{eq::lambda_2_i} also fluctuates due to atomic fluctuations, we use a moving average
\begin{equation}
\lambda_{\text{avg},i}(t) = \frac{1}{N_{\lambda,\text{avg}}} \sum_{n=1}^{N_{\lambda,\text{avg}}} \lambda_{\text{min},i}(t - n \Delta t)
\label{eq::lambda_3_i}
\end{equation}
of $N_{\lambda,\text{avg}}$ timesteps.
To prevent unnecessary fluctuations of the switching function further, we introduce a minimum step size $\Delta\lambda_\text{min}$ and neglect changes smaller than this step size in form of
\begin{equation}
\lambda_{\text{avg},\text{msz},i}(t) = \left\{
\begin{array}{l}
\lambda_{\text{avg},i}(t) \text{ for } \left|\lambda_{\text{avg},i}(t) - \lambda_{\text{avg},\text{msz},i}(t-\Delta t)\right|\\
\phantom{\lambda_{\text{avg},i}(t) \text{ for }|} \geq \Delta\lambda_\text{min} \text{ or } \lambda_{\text{avg},i}(t)\in\{0,1\}, \\
\lambda_{\text{avg},\text{msz},i}(t-\Delta t) \text{ otherwise\,.}
\end{array}
\right.
\label{eq::lambda_4_i}
\end{equation}
When an atom is detected for a precise calculation according to $\lambda_{\text{min},i}(t)$ it takes $N_{\lambda,\text{avg}}$ timesteps until $\lambda_{\text{avg},i}$ is changed completely due to the time average.
Therefore, a change of $\lambda_{\text{avg},i}$ in the order of magnitude of $1/N_{\lambda,\text{avg}}$ is expected.
Changes of $\lambda_{\text{avg},\text{msz},i}$ smaller than $1/N_{\lambda,\text{avg}}$ can be neglected without disturbing the smooth transition from the fast interatomic potential to the precise one or vice versa in a relevant way.
Thus, the use of an appropriately small $\Delta\lambda_\text{min}$ prevents only $\lambda$ changes due to atomic fluctuations.
We use $\lambda_{\text{avg},\text{msz},i}$ according to \cref{eq::lambda_4_i} as switching function for energy- and force-calculations as the CSP precisely detects defects and surface atoms and the switching function ensures precise forces on these detected atoms.

\section{Local thermostat}
\label{sec::appendix_local_thermostat}
The local thermostat needs to change the kinetic energy of particles in order to correct the energy error $\Delta E_i$ according to \cref{eq::Delta_E_i} introduced by changed switching parameters and thereby conserve the energy.

\subsection{Error calculation}
\label{sec::appendix_local_thermostat_error_calculation}
$\Delta E_i$ is the energy difference of atom $i$ at the end of a time step between the conservative reference system with constant switching parameters and the system with updated switching parameters.
Therefore, one has to finish the calculation of an integration step for both sets of switching parameters.
The quantities of the conservative reference system calculated with the constant switching parameter $\lambda_i^\text{c}(t+\Delta t)=\lambda_i(t)$ are denoted with $\dots^\text{c}$.
In contrast, the quantities calculated with the dynamically updated switching parameter $\lambda_i^\text{d}(t+\Delta t)$ are denoted with $\dots^\text{d}$.
The non-energy conserving force on an atom with the updated switching parameter under neglection of $\nabla_i\lambda_k$ in \cref{eq::force_hyb} is
\begin{equation}
F_i^\text{d} = \sum_k \left(- \lambda_k^\text{d} (\nabla_i E_k^\text{(f)}) - (1 - \lambda_k^\text{d}) (\nabla_i E_k^\text{(p)}) \right)\,.
\label{eq::F_i_d}
\end{equation}
With the same force equation like \cref{eq::F_i_d} but $\lambda_k^\text{c}$ instead of $\lambda_k^\text{d}$, we get the conservative reference force $F_i^\text{c}$.
Since the form of the equation is the same and both forces depend on the same atomic positions $x_i(t+\Delta t)$ one can easily calculate $F_i^\text{d}$ and $F_i^\text{c}$ during the force calculation by just summing up two forces weighted with the corresponding switching parameter.

To measure the violation of energy conservation (cmp. \cref{eq::Delta_E_i} by neglecting $\nabla_i\lambda_k$ and using $F_i^\text{d}$ according to \cref{eq::F_i_d}, we calculate both the potential energy difference $\Delta V_i(t + \Delta t) = E_i^\text{c}(t + \Delta t) - E_i^\text{d}(t + \Delta t)$ and the kinetic energy difference $\Delta T_i(t+\Delta t)= (|\vec{v}_i^\text{c}(t + \Delta t)|^2 - |\vec{v}_i^\text{d}(t + \Delta t)|^2)m_i / 2$ at the end of the timestep.
The potential energy difference is given as
\begin{equation}
\begin{split}
\Delta V_i(t + \Delta t)
           =& \left(\lambda_i^\text{c}(t + \Delta t) - \lambda_i^\text{d}(t + \Delta t)\right)\\
            & \times \left(E_i^\text{(f)}(t + \Delta t) -  E_i^\text{(p)}(t + \Delta t)\right)\,.
\end{split}
\label{eq::Delta_V_i}
\end{equation}
One should note, that \cref{eq::Delta_V_i} only uses already calculated potential energies and does not introduce further overhead.
The switching parameter is known for all atoms and one does not need any potential energies for atoms $i$ with $\lambda_i^\text{c}(t + \Delta t) = \lambda_i^\text{d}(t + \Delta t)$.
The kinetic energy difference is given as
\begin{equation}
\begin{split}
\Delta T_i(t+\Delta t) = \sum_{\alpha=1}^3\Bigg(&\frac{(\Delta t)^2}{8m_i} \left(\left(F_{i\alpha}^\text{c}(t + \Delta t)\right)^2 -\left(F_{i\alpha}^\text{d}(t + \Delta t)\right)^2\right)\\
&+\Big(F_{i\alpha}^\text{c}(t + \Delta t) - F_{i\alpha}^\text{d}(t + \Delta t)\Big)\\
&\phantom{+\Big(}\times v_{i\alpha}\left(t+\frac{\Delta t}{2}\right) \frac{\Delta t}{2} \Bigg)\,,\label{eq::Delta_T_i}
\end{split}
\end{equation}
whereas $\alpha$ denotes the spatial component of a vector.
It is important to use only the forces of the interatomic potential as $F_i^\text{c}$ and $F_i^\text{d}$ and to neglect any present external forces since such additional forces may not be energy- and momentum-conserving.

\subsection{Error correction}
\label{sec::appendix_local_thermostat_error_correction}
Coupling particles to a local heat bath through collisions of two particles according to Lowe-Andersen\cite{lowe_andersen_thermostat}, for example, is therefore impractical since the direction of the momentum change of the two particles is in the direction of the relative distance of both particles.
This predefined direction of a momentum change for two particles may result in very high and unphysical forces when the relative momentum of the two considered particles is not parallel to the direction of the momentum change.
Thus, one needs to apply a correction to the set $\Omega_i$ of atoms rather than to a pair of atoms to distribute the additional force over multiple atoms.
We get a minimum additional force when a particles momentum is rescaled since the direction of force and momentum change is equal in this case, but this would violate momentum conservation.
Thus, we need to rescale with respect to the center-of mass velocity
$\vec{v}_\text{cm}(\Omega_i) = (\sum_{j\in\Omega_i} m_j \vec{v}_j)/(\sum_{j\in\Omega_i} m_j)$
of the rescaled atoms.
We want to rescale only the relative momentum
$\vec{p}_{\text{rel},j}(\Omega_i) = \vec{p}_j(\Omega_i) - m_j\vec{v}_\text{cm}(\Omega_i)$
of all particles $j\in\Omega_i$ by the factor $\beta(\Omega_i)$, namely
\begin{equation}
\vec{p}_{j,\text{resc}}(\Omega_i) = m_j \vec{v}_{cm}(\Omega_i) + \beta(\Omega_i)\vec{p}_{\text{rel},j}(\Omega_i)\,.
\label{eq::p_resc_omega}
\end{equation}
The total relative momentum $\sum_{j\in\Omega_i} \vec{p}_{\text{rel},j}=\vec{0}$ vanishes and the rescaling according to \cref{eq::p_resc_omega} therefore conserves the momentum.
The kinetic energy change in $\Omega_i$ due to the rescaling is
\begin{equation}
\Delta T_\text{resc}(\Omega_i) = (\beta(\Omega_i)^2 - 1) \sum_{j\in\Omega_i} \frac{1}{2 m_j} p_{\text{rel},j}(\Omega_i)^2\,.
\end{equation}
To correct the detected energy error $\Delta E_i$ according to \cref{eq::Delta_E_i}, we need to select $\beta(\Omega_i)$ to satisfy $\Delta E_i = \Delta T_\text{resc}(\Omega_i)$, namely
\begin{equation}
\beta(\Omega_i) = \sqrt{\frac{\Delta E_i}{\sum_{j\in\Omega_i} \frac{1}{2 m_j} p_{\text{rel},j}(\Omega_i)^2} + 1}\,.
\label{eq::beta_omega_i}
\end{equation}
We neglect $\beta(\Omega_i) = -\sqrt{\dots}$ although it fulfills $\Delta E_i = \Delta T_\text{resc}(\Omega_i)$, because it alters the momentum direction for $\Delta E_i = 0$.
Rescaling according to \cref{eq::beta_omega_i} conserves the momentum for $\Delta E_i=0$.
It is important to note that the rescaling factor $\beta$ depends on the momentum before the rescaling.
Therefore, our method presents challenges in terms of parallelization as one needs a sub-domain decomposition with sub-domains which can be treated independently.
We did not develop and implement such a sub-domain decomposition as it is not the focus of our work.
Instead, we only rescale locally administered particles and no ghost particles.
We use a random selection of locally administered particles of the neighbor list of particle $i$ and the particle $i$ itself as $\Omega_i$.
Thereby, the energy correction by rescaling relative momenta can be applied per processor without communication for all particles with an energy error $\Delta E_i\neq 0$.
Note that the application of this local thermostat contains essentially two stochastic components.
Firstly, the selection of neighboring atoms that are rescaled is stochastic, namely $\Omega_i$.
Secondly, the direction of the effective force is the direction of the relative momentum $\vec{p}_{\text{rel},j}(\Omega_i)$ (cmp. \cref{eq::p_resc_omega}) and thus stochastic.

The maximum possible decrease of the relative momentum per particle is to decrease the relative momentum for all particles $j\in\Omega_i$ to $\SI{0}{\kilo\gram\metre/\second}$ with $\beta(\Omega_i)=0$.
Thus, according to \cref{eq::beta_omega_i} energy errors $\Delta E_i < - \sum_{j\in\Omega_i} p_{\text{rel},j}(\Omega_i)^2 / (2 m_j)$ cannot be corrected as they correspond to a negative radicand in \cref{eq::beta_omega_i} and imply $\beta\not\in\mathbb{R}$.
Hence, one needs to include more particles in the rescaling in this case.
Concrete prevention strategies and workarounds in LAMMPS for negative radicands are discussed in \cref{sec::negative_radicand_lammps}.
However, the occurrence of negative radicands depends on the simulation setup and the combined potentials and is by our observation an extremely rare event which does not occur at all in many simulations.

The presented rescaling approach fulfills momentum conservation as only relative momenta are rescaled and results in energy conservation for the forces according to \cref{eq::F_i_d}.
One can then use $F_i^\text{d}(t+\Delta t)$ as $F_i(t+\Delta t)$ and the rescaled velocities as $v_i(t+\Delta t)$ at the begin of the next integration step.

This local thermostat is a tool to correct locally a known energy gain or loss of an atom $i$ during the integration step by adjusting the kinetic energy in the local environment of this atom.
We exactly know the energy error due to the neglection of $\nabla_i\lambda_k$ in \cref{eq::force_hyb} and correct this error to conserve the energy within numerical precision.

\section{Handling negative radicands in LAMMPS}
\label{sec::negative_radicand_lammps}
Energy errors $\Delta E_i < - \sum_{j\in\Omega_i} p_{\text{rel},j}(\Omega_i)^2 / (2 m_j)$ cannot be corrected by the local thermostat as they correspond to a negative radicand in \cref{eq::beta_omega_i} and imply $\beta\not\in\mathbb{R}$.
Hence, one needs a larger neighbor list to randomly draw atoms as $\Omega_i$ in case of a negative radicand.
As a dynamic neighbor-list cutoff is not implemented in LAMMPS, one needs to use the last restart file to run the simulation with a larger neighbor-list cutoff.
Requesting a larger neighbor-list cutoff as precautionary measure is possible, but would reduce the performance of the whole simulation rather than just of a small subset of timesteps.
As we use a parallel-computing approach including a load-balancing scheme problems due to the domain geometry are possible.
When the load balancer reduces the size of a domain, it reduces thereby also the number of local particles which are available for rescaling.
Thus, in case of a negative radicand one can also restart the simulation with temporary disabled load balancing to avoid the existence of small domains. For details regarding the load balancing see \cref{sec::load_balancing}.

\section{Optimizing the EAM potential}
\label{sec::Mishins_EAM_potential}
\begin{table*}
\caption{\label{tab::copper_fit_300K_Mishin_parameters}Optimized parameters of the EAM Copper potential compared with the original values of EAM2 of Ref. \protect{\cite{mishin}}.}
\centering
\begin{tabular}{lrrlrr}
\hline\hline
Parameter                                & Value original & Value fit 300K & Parameter                             & Value original & Value fit 300K\\\hline
$E_1 / \si{\electronvolt}$               &  2.01458e+02 &  2.01486e+02   & $\delta / \si{\angstrom}$             & 8.62250e-03  &  2.18016e-02 \\
$E_2 / \si{\electronvolt}$               &  6.59288e-03 &  6.44057e-03   & $F_0^\text{Fit} / \si{\electronvolt}$ & 0.00000e+00  & -1.50036e-01 \\
$\xi^{(0)} / \si{\electronvolt}$           & -2.30000e+00 & -2.30089e+00   & $h / \si{\angstrom}$                  &  5.00370e-01 &  5.00370e-01 \\
$\xi^{(2)} / \si{\electronvolt}$           &  1.40000e+00 &  1.39912e+00   & $q_1 / \si{\electronvolt}$            & -1.30000e+00 & -1.29894e+00 \\
$Q_1$                                    &  4.00000e-01 &  3.99178e-01   & $q_2 / \si{\electronvolt}$            & -9.00000e-01 & -8.99672e-01 \\
$Q_2$                                    &  3.00000e-01 &  3.00455e-01   & $q_3 / \si{\electronvolt}$            &  1.80000e+00 &  1.79940e+00 \\
$S_1 / \si{\electronvolt\angstrom^{-4}}$ &  4.00000e+00 &  4.00000e+00   & $q_4 / \si{\electronvolt}$            &  3.00000e+00 &  2.99944e+00 \\
$S_2 / \si{\electronvolt\angstrom^{-4}}$ &  4.00000e+01 &  4.00000e+01   & $r_0^{(1)} / \si{\angstrom}$          &  8.35910e-01 &  8.36347e-01 \\
$S_3 / \si{\electronvolt\angstrom^{-4}}$ &  1.15000e+03 &  1.15000e+03   & $r_0^{(2)} / \si{\angstrom}$          &  4.46867e+00 &  4.46690e+00 \\
$a$                                      &  3.80362e+00 &  3.80541e+00   & $r_0^{(3)} / \si{\angstrom}$          & -2.19885e+00 & -2.19834e+00 \\
$\alpha_1 / \si{\angstrom^{-1}}$         &  2.97758e+00 &  2.96972e+00   & $r_0^{(4)} / \si{\angstrom}$          & -2.61984e+02 & -2.61984e+02 \\
$\alpha_2 / \si{\angstrom^{-1}}$         &  1.54927e+00 &  1.55510e+00   & $r_\text{s}^{(1)} / \si{\angstrom}$   &  2.24000e+00 &  2.24000e+00 \\
$\beta_1 / \si{\angstrom^{-2}}$          &  1.73940e-01 &  1.66589e-01   & $r_\text{s}^{(2)} / \si{\angstrom}$   &  1.80000e+00 &  1.80000e+00 \\
$\beta_2 / \si{\angstrom^{-1}}$          &  5.35661e+02 &  5.35661e+02   & $r_\text{s}^{(3)} / \si{\angstrom}$   &  1.20000e+00 &  1.20000e+00 \\
$r_\text{c} / \si{\angstrom}$            &  5.50679e+00 &  5.50679e+00
\\\hline\hline
\end{tabular}
\end{table*}
A EAM potential is given according to \cref{eq::energy_eam} by an embedding function $\xi$, an electron density $\zeta$ and a pair potential $\Phi$.
We use a Copper potential from Ref. \cite{mishin} which is given by the pair potential
\begin{equation}
\begin{split}
\Phi(x) =& \Big( E_1 M(x,r_0^{(1)},\alpha_1) +E_2 M(x,r_0^{(2)},\alpha_2) + \delta\Big)\\
&\times \psi\left(\frac{x-r_\text{c}}{h}\right) - \sum_{n=1}^{3} \Theta(r_\text{s}^{(n)}-x) S_n (r_\text{s}{(n)} -x)^4\,,
\end{split}
\end{equation}
where
\begin{equation}
\psi(x) = x^4 / (1 + x^4) \Theta(-x)
\end{equation}
and
\begin{equation}
M(x,r_0,\alpha) = \text{exp}(-2\alpha(x-r_0)) -2 \text{exp}(-\alpha(x-r_0))
\end{equation}
with the Heaviside step function $\Theta(x) = 1 \text{ for } x \geq 0, 0 \text{ for } x < 0$.
The embedding function is
\begin{equation}
\xi(x) =
\left\{
\begin{array}{ll}
\xi^{(0)} + \frac{1}{2} \xi^{(2)} (x-1)^2 & \text{if } x \le 1 \\
\phantom{\xi^{(0)}} + \sum_{n=1}^4 q_n (x-1)^{n+2} \\
\Big(\xi^{(0)} + \frac{1}{2} \xi^{(2)} (x-1)^2 + q_1 (x-1)^{3} & \text{otherwise}\,.\\
\phantom{\Big(}+ Q_1 (x-1)^4\Big)/(1+Q_2(x -1)^3)
\end{array}
\right.
\label{eq::mishin_embedding_function}
\end{equation}
The electron density is
\begin{equation}
\begin{split}
\zeta(x) =& \left( a\ \text{exp}\left(-\beta_1 (x-r_0^{(3)})^2\right) + \text{exp}\left(-\beta_2(x-r_0^{(4)})\right) \right)\\
& \times \psi\left((x-r_\text{c})/h\right)\,.
\end{split}
\end{equation}
The EAM potential is optimized as described in \cref{sec::eam_ace} with atomicrex.
The target values and tolerances for the minimized loss function (\ref{eq::atomicrex_loss_function}) are shown in \cref{tab::target_values_copper}.
The fitted parameters are shown in \cref{tab::copper_fit_300K_Mishin_parameters}.

\section{Parameter selection for an adaptive-precision potential}
\label{sec::parameter_selection}
The parameters of the hybrid potential are listed in \cref{tab::model_parameters} and will be explained in more detail in the following.
\begin{table}
\caption{\label{tab::model_parameters}Parameters of the adaptive-precision model and values of the parameter set Hyb1 for copper at $\SI{300}{\kelvin}$.}
\begin{center}
\begin{tabular}{llllll}
\hline\hline
parameter & value & parameter & value & parameter & value \\
\hline
$N_\text{buffer}$          & 0\,atoms\phantom{\quad} & $\text{CSP}_\text{hi}$     & $\SI{3.0}{\angstrom^2}$             & $N_{\lambda,\text{avg}}$   & 110                          \\
$N_{\text{CSP},\text{avg}}$& 110                     & $r_{\lambda,\text{lo}}$    & \SI{4.0}{\angstrom}                 & $\Delta\lambda_\text{min}$ & $1.0/N_{\lambda,\text{avg}}$ \\
$\text{CSP}_\text{lo}$     & $\SI{2.5}{\angstrom^2}$ & $r_{\lambda,\text{hi}}$    & \SI{12.0}{\angstrom}\phantom{\quad} & $|\Omega_i|$               & 800\,atoms                   \\
\hline\hline
\end{tabular}
\end{center}
\end{table}

\subsection{Centro-symmetry parameter}
$N_\text{buffer}$ is used to ensure all relevant neighboring atoms are used in the calculation of the centro-symmetry parameter according to \cref{eq::csp_1_i}.
For copper, we have not observed the case of unexpected high centro-symmetry parameters we discussed in \cref{Fig::csp_calculation_problem_bcc_tungsten} for tungsten.
Therefore, we use $N_\text{buffer}=0$.
An unexpected high CSP is less likely for copper as the distance in lattice constants between the first and second neighbor shell is for BCC with $1-\sqrt{3/4}\approx0.13$ smaller than for FCC with $1-\sqrt{1/2}\approx0.29$.

\begin{figure}[]
\begin{center}
\includegraphics[width=3.30in]{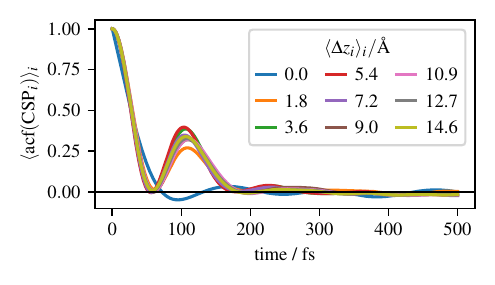}
\end{center}
\caption{\label{Fig::Copper_acf}
Mean autocorrelation functions of the centro-symmetry parameter $\text{CSP}_{i}$ according to \cref{eq::csp_1_i}.
Each shown graph is the mean of the autocorrelation functions (acf) of one atom layer in $z$-direction.
The mean distance $\langle\Delta z_i\rangle_i$ of the corresponding atom layer to the (100)-copper surface is given in the legend.
The autocorrelation functions are calculated from data of a $\SI{5}{\pico\second}$ simulation with an adaptive-precision potential.
}
\end{figure}

The centro-symmetry parameter of an atom $i$ fluctuates due to atomic fluctuations.
To suppress this fluctuations, we use a time average of $N_{\text{CSP},\text{avg}}$ timesteps in \cref{eq::csp_2_i}.
To determine $N_{\text{CSP},\text{avg}}$, we measured autocorrelation functions (acf) for $\SI{5}{\pico\second}$ near a (100)-copper surface as shown in \cref{Fig::Copper_acf} as mean per (100) atom layer dependent on the distance to the surface.
The acf of the CSP has a peak at about $\SI{110}{\femto\second}$ for all layers apart from the surface layer.
As the CSP depends on the local environment, a different acf for the surface atoms is expected.
Since the time averaging of the centro-symmetry parameter is used to average out thermal fluctuations, we want to average the CSP up to the peak of the acf of non-surface atoms at $\SI{110}{\femto\second}$.
With the used timestep $\Delta t=\SI{1}{\femto\second}$ follows $N_{\text{CSP},\text{avg}} = 110$.
\begin{figure*}[]
\begin{center}
\includegraphics[width=6.69in]{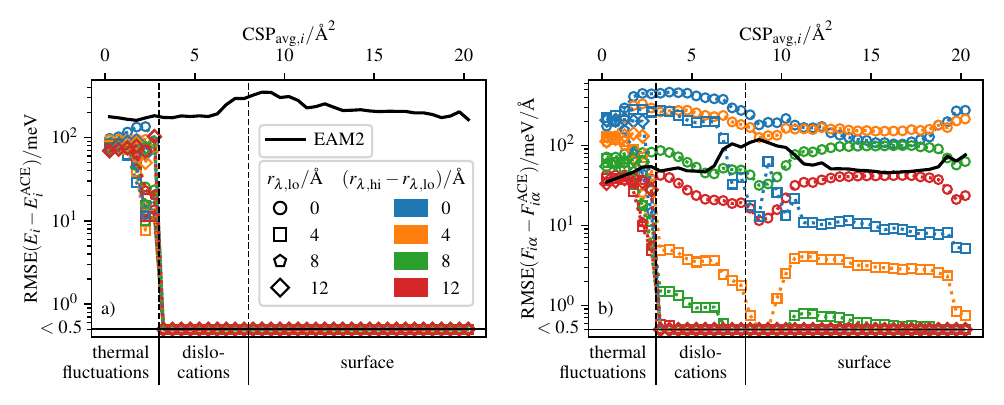}
\end{center}
\caption{\label{Fig::Copper_switching_zone_width}
Root mean square error (RMSE) of the adaptive-precision potential compared with the precise ACE potential for a snapshot of a nanoindentation dependent on $r_{\lambda,\text{lo}}$ and $r_{\lambda,\text{hi}}$.
All atoms whose forces are influenced by both EAM and ACE are grouped dependent on $\text{CSP}_{\text{avg},i}$ and the RMSE is calculated per group.
Atoms with pure EAM or ACE force-contributions are not considered since their energies and forces are good.
The RMSE of ACE compared to EAM2 of \protect{\cite{mishin}} is shown as reference.
The cutoff $r_{\lambda,\text{lo}}$ is represented by different symbols and the width $r_{\lambda,\text{hi}}-r_{\lambda,\text{lo}}$ of the switching zone is color-coded.
}
\end{figure*}

The distribution of the centro-symmetry parameter $\text{CSP}_{i}$ in the training data is used to set the thresholds $\text{CSP}_\text{lo}$ and $\text{CSP}_\text{hi}$ for \cref{eq::lambda_1_i}.
The fast potential is optimized for usage at a target temperature with MD data of the precise and expensive potential.
The training data contain only thermal fluctuations of atoms with observed $\text{CSP}_i\in[\SI{0}{\angstrom^2},\SI{3}{\angstrom^2}]$.
Therefore, we distinguish between a small CSP of the thermal fluctuations and all other atoms with a higher CSP.
As atoms with $\text{CSP}_{\text{avg},i} \geq \text{CSP}_\text{hi}$ are detected for a precise calculation according to \cref{eq::lambda_1_i}, we select $\text{CSP}_\text{hi} = \SI{3}{\angstrom^2}$.
Thereby, we do not detect atoms with expected thermal fluctuations but all other atoms.
We use $\text{CSP}_\text{lo} = \SI{2.5}{\angstrom^2}$ since the difference of $\SI{0.5}{\angstrom^2}$ between the thresholds allows a smooth transition for atoms from the fast to the precise potential.
One might use a slightly different value $\text{CSP}_\text{lo}$, but far lower values limit the possible performance gain since thermally fluctuating atoms are treated partially with the precise potential.
It is essential to treat the majority of the thermally fluctuating atoms completely with EAM as one can only save compute time when ACE is not evaluated for most atoms at all.

\subsection{Switching function}
The cutoff radii $r_{\lambda,\text{lo}}$ and $r_{\lambda,\text{hi}}$ are used in the calculation of $\lambda_{\text{min},i}$ according to \cref{eq::lambda_2_i} to decrease the switching parameters for neighboring atoms of atoms, which require a precise calculation.
The force on an atom depends on all switching parameters within the force cutoff.
Thus, $r_{\lambda,\text{lo}}$ and $r_{\lambda,\text{hi}}$ increase the force precision for precisely calculated atoms.
To determine the cutoff radii $r_{\lambda,\text{lo}}$ and $r_{\lambda,\text{hi}}$, we calculate energies and forces according to \cref{eq::energy_hyb,eq::F_i_d} for one timestep with $\lambda_{\text{min},i}$ as switching function under neglection of $(\nabla_i\lambda_k)$ as discussed in \cref{sec::integration_of_motion}.
However, we do not apply a local thermostat to prevent random influence in this parameter study and calculate only one timestep.
We use a snapshot of a nanoindentation including dislocations.
The energy error compared to the precise ACE energy as shown in \cref{Fig::Copper_switching_zone_width}a vanishes by design for precisely calculated atoms independently of the varied parameters since it depends only on $\lambda_{\text{min},i}$ of the corresponding precisely calculated atom $i$.
The force error compared to the precise ACE forces is shown in \cref{Fig::Copper_switching_zone_width}b and depends for the case of thermal fluctuations mainly on the width $r_{\lambda,\text{hi}}-r_{\lambda,\text{lo}}$ of the switching zone between precise and fast atoms but is independent of $r_{\lambda,\text{lo}}$.
The force error for thermal fluctuations can be larger than for the EAM reference force since only the total forces rather than the force contributions of EAM and ACE are fitted as shown in \cref{Fig::copper_f_ij_comparison}.
Therefore, a switching zone is needed to change smoothly between EAM and ACE atoms.
The larger the switching zone, the smaller the force error on thermal fluctuations, but the more computationally expensive the calculation becomes since more precise calculations are required.
We selected a switching zone width of $r_{\lambda,\text{hi}}-r_{\lambda,\text{lo}}=\SI{8}{\angstrom}$.
The force error for precisely calculated atoms depends primarily on $r_{\lambda,\text{lo}}$.
We selected $r_{\lambda,\text{lo}}=\SI{4}{\angstrom}$.

The switching parameter $\lambda_{\text{min},i}$ according to \cref{eq::lambda_2_i} of atom $i$ fluctuates as it depends on the distance $r_{ij}$ to neighboring atoms $j$, which require a precise calculation.
Thus, we use a time average of $N_{\lambda,\text{avg}}$ timesteps in \cref{eq::lambda_3_i} to average out these fluctuations of the switching parameter.
As this is the same motivation like for the time averaging of the centro-symmetry parameter in \cref{eq::csp_2_i}, we use the same number $N_{\lambda,\text{avg}} = N_{\text{CSP},\text{avg}} = 110$ of averaged timesteps.

\subsection{Local thermostat}
\begin{figure}[]
\begin{center}
\includegraphics[width=3.37in]{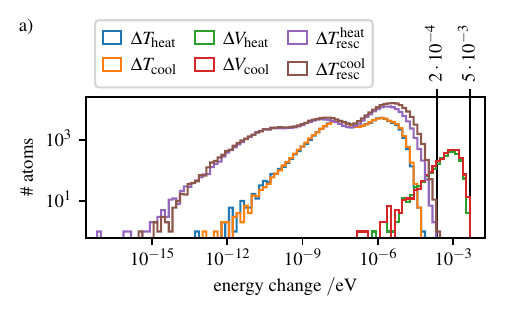}
\includegraphics[width=3.37in]{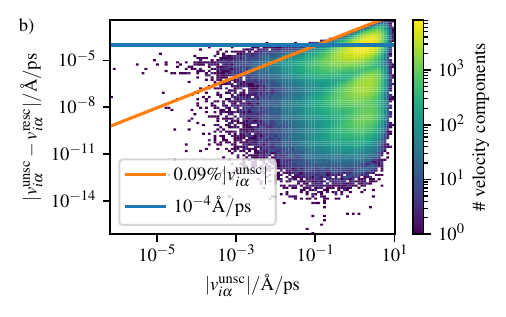}
\includegraphics[width=3.37in]{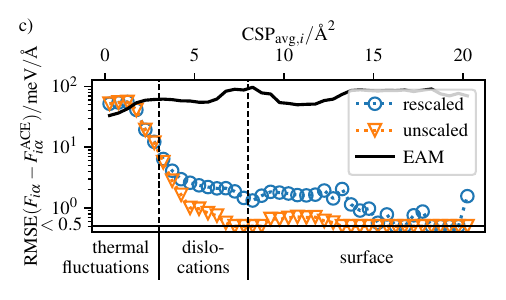}
\end{center}
\caption{\label{Fig::influence_rescaling_snapshot}
Rescaling-related quantities of a nanoindentation at $\SI{300}{\kelvin}$ and $\SI{20}{\angstrom}$ indentation depth.
a) Histogram of potential energy change $\Delta V_i$ according to \cref{eq::Delta_V_i}, kinetic energy change $\Delta T_i$ according to \cref{eq::Delta_T_i} and kinetic energy change $\Delta T_\text{resc}$ due to the applied rescaling according to \cref{eq::p_resc_omega} separated for positive values denoted with 'heat' and negative values denoted with 'cool'.
b) 2D-Histogram of the velocity $|v_{i\alpha}^\text{unsc}|$ before the rescaling step compared with the velocity change $|v_{i\alpha}^\text{unsc}-v_{i\alpha}^\text{resc}|$ during the rescaling step, where $i$ denotes a particle and $\alpha$ a spatial dimension.
c) Root mean square error (RMSE) of different forces compared with the precise ACE potential.
All rescaled atoms are grouped dependent on $\text{CSP}_{\text{avg},i}$ and the RMSE in calculated per group.
The applied forces according to \cref{eq::F_i_d} are denoted with 'unscaled'.
The graph denoted with 'rescaled' corresponds to the applied forces plus theoretical forces according to \cref{eq::additional_force_contribution} due to the rescaling.
The forces of the optimized EAM potential used for the adaptive-precision potential are denoted with 'EAM' and given as reference.
}
\end{figure}
An energy error $\Delta E_i$ according to \cref{eq::Delta_E_i} of atom $i$ is corrected by rescaling momenta of the set of random neighboring atoms $\Omega_i$ according to \cref{eq::beta_omega_i}.
The distribution of $\Delta E_i$ in \cref{Fig::influence_rescaling_snapshot}a for a snapshot of a copper nanoindentation at $\SI{20}{\angstrom}$ indentation depth and $\SI{300}{\kelvin}$ shows that energy corrections need to be applied up to an order of magnitude of $\SI{1}{\milli\electronvolt}$ due to changes of the potential energy.
The energy difference between ACE and EAM is in the order of $100\si{\milli\electronvolt}$ according to \cref{Fig::Copper_switching_zone_width}a and the minimum change of the switching parameter is $\Delta\lambda_\text{min}=1/N_{\lambda,\text{avg}}\approx{10^{-2}}$.
Thus, the energy fluctuations due to the change of $\lambda_i$ in the order of $\SI{1}{\milli\electronvolt}$ are expected.
The kinetic energy $T_i$ of an atom at $\SI{300}{\kelvin}$ is $T_i=(3/2)k_\text{B}T\approx{\SI{39}{\milli\electronvolt}}$, where $k_\text{B}$ is the Boltzmann constant and $T$ the temperature.
The kinetic energy changes only slightly with an appropriate timestep and the influence of a $\lambda$ update on the forces and therefore on the kinetic energy change is even lower.
Therefore, the contribution of the kinetic energy to the energy error is magnitudes smaller than the potential-energy contribution as shown in \cref{Fig::influence_rescaling_snapshot}a.
Nevertheless, a kinetic energy change due to rescaling in the order of $\SI{1}{\milli\electronvolt}$ is not negligible compared to an average kinetic energy of $\SI{39}{\milli\electronvolt}$ at $\SI{300}{\kelvin}$.
Hence, we need to distribute the effect of rescaling onto a larger number of particles in order to minimize effects on single atoms.
In simulations we use a maximum number $|\Omega_i|=800$ of rescaled neighbors.
The velocity change due to rescaling is shown in \cref{Fig::influence_rescaling_snapshot}b for the snapshot of the nanoindentation.
The absolute velocity change $|v_{i\alpha}^\text{unsc}-v_{i\alpha}^\text{resc}|$ is smaller than $0.1\% v_{i\alpha}^\text{unsc}$ for $v_{i\alpha}^\text{unsc} \geq \SI{0.1}{\angstrom/\pico\second}$ and smaller than $\SI{0.1}{\milli\angstrom/\pico\second}$ for about $v_{i\alpha}^\text{unsc} \leq \SI{0.1}{\angstrom/\pico\second}$.
Thus, the velocity changes are small but nevertheless may influence the dynamics of the system.
To characterize this effect further, we translate the velocity changes into effective forces.
Velocities are updated by the integrator according to $\Delta v = (F/m)\Delta t$(cmp. \cref{eq::vv_v1,eq::vv_v2}).
Hence, the additional force contribution is given as
\begin{equation}
\Delta F_{i\alpha} = (v_{i\alpha}^\text{unsc}-v_{i\alpha}^\text{resc}) m_i / \Delta t
\label{eq::additional_force_contribution}
\end{equation}
for all particles $i$ and spatial dimensions $\alpha$.
The force $\Delta F_{i\alpha}$ according to \cref{eq::additional_force_contribution}, however, is a result of the neglection of $\nabla_i\lambda_k$ in \cref{eq::force_hyb} as discussed in \cref{sec::integration_of_motion}.
The velocity updates in \cref{Fig::influence_rescaling_snapshot}b correspond to the theoretical forces in \cref{Fig::influence_rescaling_snapshot}c.
These forces, including rescaling, on precisely calculated atoms are within a tolerance up to $\SI{10}{\milli\electronvolt/\angstrom}$ compared to the precise ACE forces.
The potential energy $E_i$  depends, according to \cref{eq::energy_hyb}, only on the switching parameter $\lambda_i$ of atom $i$.
Therefore, the precision of the potential energies $E_i$ is unaffected by the rescaling.

\section{Atom subgroups for load balancing}
\label{sec::atom_subgroups_load_balancing}
We execute firstly the calculation of the fast potential (FP), secondly the precise potential (PP), thirdly $\text{CSP}_{i}$ (CSP) and fourthly $\lambda_{\text{min},i}$ from $\lambda_{0,j}$ ($\lambda$).
Which potential needs to be calculated depends on the switching parameter $\lambda$.
The centro-symmetry parameter is calculated for the group of atoms with a changeable switching parameter.
The CSP is not calculated for atoms with a constant switching parameter which is useful for static atoms used at an open boundary.
The fourth subroutine is decreasing, if possible, the switching parameter $\lambda_{\text{min},i}$ compared to $\lambda_{0,i}$ for neighboring atoms of atoms, which need to be calculated precisely, by calculating $\lambda_{\text{min},i}$ according to \cref{eq::lambda_2_i}.
Atoms $i$ with $\lambda_{0,i}=1$, which do not require a precise calculation, do not influence $\lambda_{\text{min},i}$ of neighboring atoms.
As $\lambda_{0,i}=1$ should apply for most of the atoms, these atoms should not cause load for the calculation of $\lambda_{\text{min},i}$, which is achieved by iterating over the neighboring atoms of atoms with $\lambda_{0,i}\in[0,1)$.

\section{Adjusting potential calculation for load balancing}
\label{sec::adjustments_for_load_balancing}
For all four force subroutines to be executed independently one after the other on all processors, it is important that none of the subroutines contains communication with other processors as communication is a synchronization point within a force-calculation subroutine and may include high additional waiting times.
Thus, avoiding communication during the force-calculation subroutines is essential to allow effective load balancing.
The EAM calculation in LAMMPS includes communication of the derivative of the embedding function and the electron density within the force-calculation routine.
In order to avoid communication during the force calculation subroutines, we compute the corresponding quantities on all processors which require them.
This results is double calculations, but is acceptable in our case to achieve proper load balancing.
The ACE calculation did not require any adjustments due to communication.

\section{Processor dependency of subprocess work}
\label{sec::processor_dependency_work}
\begin{figure}
\includegraphics[width=3.34in]{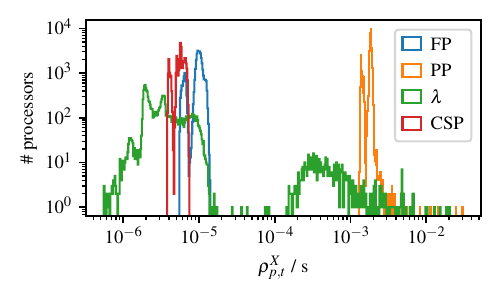}
\caption{\label{Fig::load_balancing_work_distribution}
Histogram of the average work $\rho_{p,t}^{(\mathcal{X})}$ for all load-balancing steps of a nanoindentation in a cubic system of $100^3$ unit cells and load balancing with a staggered grid domain decomposition.
}
\end{figure}
\begin{figure}[]
\includegraphics[width=3.34in]{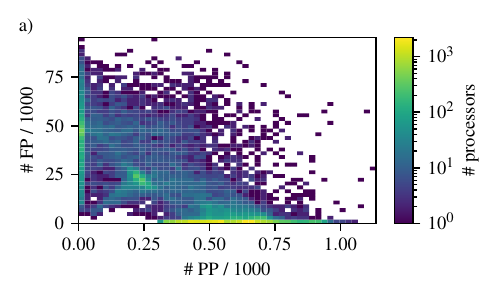}
\includegraphics[width=3.37in]{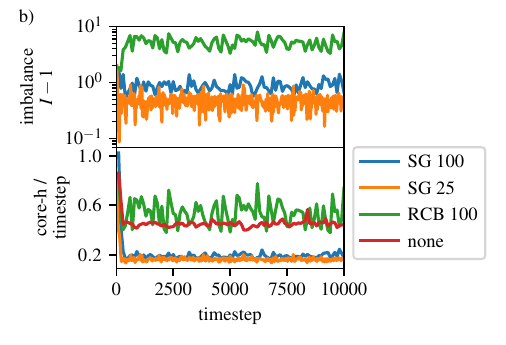}
\includegraphics[width=3.37in]{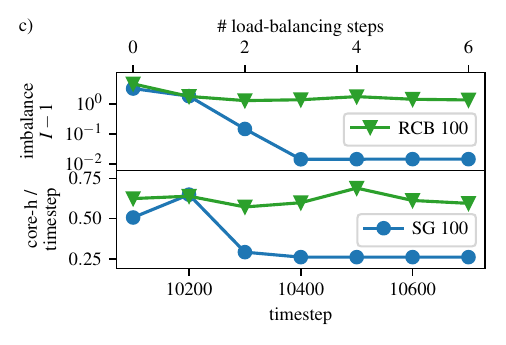}
\caption{\label{Fig::load_balancing_dynamic}
a) Number of atoms per processor which require the calculation of the fast potential (FP) and the precise potential (PP) for all load-balancing steps of a nanoindentation in a cubic system of $100^3$ unit cells and load balancing with a staggered grid domain decomposition.
b) Imbalance $I$ according to \cref{eq::imbalance} and core-h per timestep during equilibration of a surface in a NVT ensemble compared for different load-balancing methods.
The staggered grid method (SG) uses per processor $p$ the work $\rho_p^{(\mathcal{X})}$ of all four force subroutines $\mathcal{X}$, whereas the recursive coordinate bisectioning (RCB) method only uses one work $\rho_p$ per processor and is thus worse balanced.
The number behind the load balancing method in the legend is the number of timesteps after which the corresponding load balancer is called.
c) Load balancing in the system of c) but with static atoms only to test the behavior of the load balancers.
}
\end{figure}
Histograms of the work $\rho_p^{(\mathcal{X})}$ according to \cref{eq::rho_p_X} used at all load balancing steps of a nanoindentation calculated on 384 processors on JURECA-DC\cite{jureca} are shown in \cref{Fig::load_balancing_work_distribution}.
$\rho_{p,t}^\lambda$ with $\langle\rho_{p,t}^\lambda\rangle_{p,t} = \SI{48.6}{\micro\second}$ is distributed over four orders of magnitude since the calculation time required on atom $i$ depends also on $\lambda_{0,j}$ of neighboring atoms.
Since the number of atoms with $\lambda_{0,i}\neq1$ should be small compared with the number of all atoms, we iterate over all atoms with $\lambda_{0,i}\neq1$ and calculate $\lambda_{\text{min},j}$ for all neighbors $j$ of $i$.
As we search a minimum switching parameter $\lambda_{\text{min},j}$ per neighbor, we can abort the calculation for $\lambda_{\text{min},j}\leq\lambda_{0,i}\neq1$ without calculating $r_{ij}$ and $f^{(cut)}$.
As the calculation of $\lambda_{\text{min},j}$ is only required on the processor which administrates particle $j$ it is not required to compute the parameter $\lambda_{\text{min}}$ for ghost particles.
Thus, the work caused by an atom $i$ depends on the neighboring atoms and is not constant for all processors.
The histograms show $\langle\rho_{p,t}^\text{FP}\rangle_{p,t} = \SI{9.6}{\micro\second}$, $\langle\rho_{p,t}^\text{PP}\rangle_{p,t} = \SI{1.7}{\milli\second}$ and $\langle\rho_{p,t}^\text{CSP}\rangle_{p,t} = \SI{5.6}{\micro\second}$, whereas there are two peaks next to each other since the calculation times are dependent on the number of neighbors which is about a factor of two smaller for surface atoms.
Therefore, one cannot use one constant processor-independent time per atom.
The histograms show the need to measure the required work per atom per processor.
Furthermore, the histograms show that the fraction of surface atoms is non-negligible.
The atoms at the surface are detected by the centro-symmetry parameter and thus require a precise calculation and hence the domains administered by a processor are small.
Note that excluding surface atoms from the precise calculation is possible by comparing the CSP to a given reference configuration, for example initial or equilibrium configuration, instead of using the absolute value of the CSP as input for the switching function in \cref{eq::lambda_1_i}.

\section{Dynamic-load balancing details}
\label{sec::dynamic_load_balancing_details}
We use a nanoindentation to discuss the challenges of dynamic load-balancing of a simulation with an adaptive-precision potential in the following.
The numbers of precise and fast calculations per processor are shown in \cref{Fig::load_balancing_dynamic}a as 2D-histogram of all load-balancing steps, whereas load-balancing is done as described in \cref{sec::load_balancing}.
The 2D-histogram shows that balancing few precise calculations on some processors with much more fast calculations on other processors works.
More precisely, the domains with the fewest fast calculations perform on average 568 precise calculations, whereas the domains with the fewest precise calculations perform on average 44245 fast calculations.

\begin{figure*}[]
\begin{center}
\includegraphics[width=6.69in]{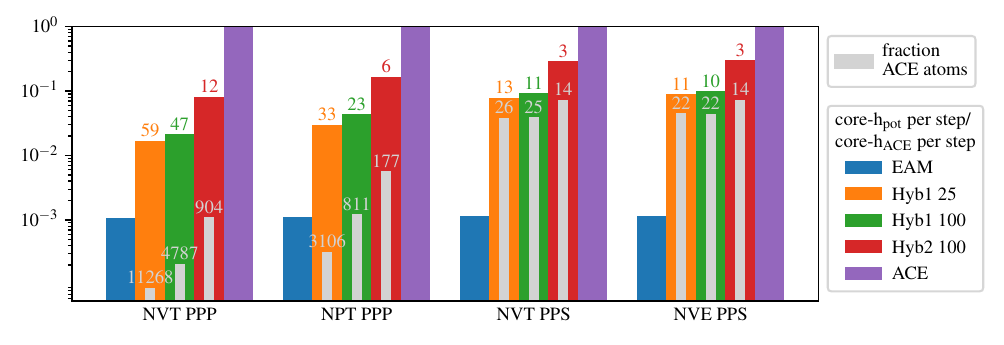}
\end{center}
\caption{\label{Fig::speedup_vs_ace}
Total computation time of nanoindentations with 4 million atoms ($100^3$ unit cells) simulated with adaptive-precision potentials compared to ACE and EAM nanoindentations.
The number behind the potential in the legend is the number of timesteps after which the staggered-grid load balancer is called.
EAM, hybrid and ACE simulations are calculated on 128, 384 and 2048 cores of JURECA-DC respectively.
The fraction of used core-h compared to ACE is given on top of the bars.
The approximated fraction of ACE atoms is given in gray to visualize the overhead induced by the method.
The times are given for all four parts of a simulation separately.
NVT and NPT ensembles are simulated with periodic boundaries (PPP) to prepare a surface (PPS).
The surface is equilibrated in an NVT ensemble.
The nanoindentation itself is calculated in an NVE ensemble.
}
\end{figure*}

To analyze dynamic load balancing further and include also non-performant load-balancing strategies, we only use the start of the equilibration instead of a whole nanoindentation.
The required core-hours per timestep during the equilibration of a surface in a NVT ensemble and the imbalance are shown in \cref{Fig::load_balancing_dynamic}b for different load-balancing strategies.
Our load-balancing method applying a staggered grid (SG) and estimating the work per particle with the four force-subroutine times $\rho_p^{(\mathcal{X})}$ (cmp. \cref{eq::rho_p_X}) results in an imbalance of 1.88 when used every 100 timesteps.
Not estimating the work per particle but per processor $p$ with only one time measurement of the total force-calculation time $\tau_p$ means that the actual load distribution on the processor is unknown to the load balancer.
Thus, load balancing with only one average work $\rho_p$ (cmp. \cref{eq::rho_p_X}) per processor using the recursive coordinate bisectioning (RCB) method of LAMMPS results in an imbalance of 6.17.
The imbalance difference between SG and RCB demonstrates the benefit of using one average work $\rho_p^{(\mathcal{X})}$ per individual force subroutine $\mathcal{X}$.
The imbalance 1 of a perfectly balanced system is not reached since the system is dynamic and changes the workload in each timestep.

An atom $i$ a bulk of only thermally fluctuating atoms can change its value $\lambda_{0,i}$ to a value smaller 1 due to a spontaneous fluctuation which might be reversed in one of the next timesteps and therefore produces workload fluctuations due to changes of the neighboring $\lambda_{\text{min},j}$ parameters according to \cref{eq::lambda_2_i}.
More concrete, the atom changes the switching parameter of all neighboring atoms within the cutoff $r_{\lambda,\text{hi}}$ to a value smaller 1 which implies a precise and expensive ACE calculation.
Within the cutoff $r_{\lambda,\text{hi}}^\text{Hby1}=\SI{12}{\angstrom}$ are 626 atoms at $\SI{0}{\kelvin}$ which is more than the 568 precise calculations for the processors with the fewest fast calculations in \cref{Fig::load_balancing_dynamic}a.
Thus, one atom with $\lambda_{\text{min},i}<1$ can double the work of its processor and one cannot expect an imbalance of 1 according to \cref{eq::imbalance} for such a dynamic system.
However, the staggered-grid load balancer reaches with 1.47 a better imbalance when called every 25 timesteps since it can better follow the dynamics of the system.
When we freeze all atoms in the NVT equilibration of the surface, we get a static force calculation after $N_{\lambda,\text{avg}}^\text{Hyb1}+N_{\text{CSP},\text{avg}}^\text{Hyb1}=220$\, timesteps.
The SG method reduces the imbalance of this static system up to $1.02$ as shown in \cref{Fig::load_balancing_dynamic}c.
Hence, the staggered-grid-load balancer works, but the system dynamics is challenging.
In contrast, the RCB-load balancer cannot balance the static system due to the missing per atom and subroutine $\mathcal{X}$ force-calculation times $\rho_p^{(\mathcal{X})}$ (cmp. \cref{eq::rho_p_X}).

\section{Simulation dependency of the saved computation time}
\label{sec::speedup_equilibration}
A nanoindentation with 4 million atoms calculated on JURECA-DC\cite{jureca} with the adaptive-precision potential Hyb1 (cmp. \cref{tab::model_parameters}) requires 11 and 13 times less core-h for equilibration and nanoindentation itself like visualized in \cref{Fig::speedup_vs_ace}.
The amount of saved computation time depends on the system itself.
Dislocations develop during a nanoindentation and therefore there are more precisely calculated atoms at the end of a nanoindentation than at the begin or during the equilibration of the surface.
Hence, one requires 59 and 33 times less core-h for the simulations with periodic boundaries in all spatial directions since there are no precisely calculated atoms at the surface.
One saves even computation time for all simulation parts for Hyb2 where we increased the cutoffs $r_{\lambda,\text{lo}}=\SI{8}{\angstrom}$ and $r_{\lambda,\text{hi}}=\SI{20}{\angstrom}$ and used the remaining parameters of Hyb1.

\end{document}